\documentclass[twoside,twocolumn,english,aps,showpacs,prl,superscriptaddress]{revtex4-1}
\usepackage[T1]{fontenc}
\usepackage[latin9]{inputenc}
\usepackage{xcolor} 
\usepackage{bm}
\usepackage{amsmath}
\usepackage{amssymb}
\usepackage{graphicx}
\usepackage{esint}
\usepackage{mathrsfs}
\usepackage{booktabs}
\usepackage[colorlinks=true,allcolors=blue]{hyperref} 

\usepackage{dcolumn}   

\makeatletter

\makeatletter
\newcommand*{\rom}[1]{\expandafter\@slowromancap\romannumeral #1@}
\makeatother

\usepackage{tabularx} 
\newcolumntype{Y}{>{\centering\arraybackslash}X} 
\newcolumntype{b}{>{\hsize=1.2\hsize}X} 
\newcommand{\heading}[1]{\multicolumn{1}{c}{#1}} 

\usepackage{multirow} 

\makeatother

\usepackage{babel}
\usepackage[normalem]{ulem}
\begin{document}

\title{Elementary models of 3D topological insulators with chiral symmetry}

\author{Donghao Liu}
\affiliation{Department of Physics, Bar-Ilan University, Ramat Gan, 52900, Israel}
\author{Polina Matveeva}
\affiliation{Department of Physics, Bar-Ilan University, Ramat Gan, 52900, Israel}
\author{Dmitri Gutman}
\affiliation{Department of Physics, Bar-Ilan University, Ramat Gan, 52900, Israel}
\author{Sam T.~Carr}
\affiliation{School of Physics and Astronomy, University of Kent, Canterbury CT2 7NH, United Kingdom}

\date{\today}

\begin{abstract}
We construct a set of lattice models of non-interacting topological insulators with chiral symmetry in three dimensions.  We build a  model of the topological insulators in the class AIII by coupling lower dimensional models of $\mathbb{Z}$ classes.  By coupling the two AIII models related by time-reversal symmetry we construct  other chiral symmetric topological insulators that may also possess additional symmetries (the time-reversal and/or particle-hole). 

There are two different chiral symmetry operators for the coupled model,   that correspond to two distinct ways of defining the sublattices.  The integer topological invariant (the winding number) in case of weak coupling can be either the sum or difference of indices of the basic building blocks,  dependent on the preserved chiral symmetry operator.   The value of the topological index in case of weak coupling is determined by the chiral symmetry only and does not depend on the presence of other symmetries.  For $\mathbb{Z}$ topological classes AIII, DIII, and CI with chiral symmetry are topologically equivalent,  it implies that a smooth transition between the classes can be achieved if it connects the topological sectors with the same winding number.  We demonstrate this explicitly by proving that the gapless surface states remain robust in $\mathbb{Z}$ classes as long as the chiral symmetry is preserved, and the coupling does not close the gap in the bulk. By studying the surface states   in   $\mathbb{Z}_2$ topological classes,  we show that class CII and AII  are distinct,  and can not be adiabatically connected.

\end{abstract}

\maketitle

\section{Introduction}

While conventional phases of matter can be classified by the conventional Landau symmetry-breaking paradigm \cite{landau_statistics},  topological phase go beyond this framework.
They refer to states of matter that in addition to symmetries are characterized by their topological properties, which are robust and non-local.
A well-known example  is the integer Quantum Hall Effect  (IQHE) \cite{klitzing1980new}, 
which is observed in a two-dimensional electron gas at low temperatures and high magnetic fields. 
The transitions between states of the IQHE sample with different filling fractions are accompanied by a change in the number of edge modes and occur without any change in symmetry.  
The edge modes in IQHE are robust and are protected by the non-zero first Chern number that determines the quantized Hall conductance.  
The IQHE represents an example of a topological insulator that cannot be smoothly 
connected to a trivial insulator without closing a gap.  
By now there are many experimentally realized topological insulators \cite{hasan2010colloquium,qi2011topological,chiu2016classification,Wang2017,Kou2017}. 
The prototypical example of 1D topological insulator  is the polyacetylene polymer chain, accurately 
described   Su-Schrieffer-Heeger (SSH) model ~\cite{SSHmodel,RyuHatsugai2006, Leder2016,Bloch2013}. The latter has a  $\mathbb{Z}$ topological invariant. In  three-dimension  there are  numerous realizations of $\mathbb{Z}_2$ topological insulators \cite{fu2007topological,qi2008topological,hsieh2008topological,hsieh2009tunable,xia2009observation}.

From the mathematical perspective, the entire classification of non-interacting topological insulators is complete \cite{zirnbauer1996riemannian,altland1997nonstandard}. 
It is fully determined by dimensionality and the presence or absence of chiral symmetry  ($C$),   time-reversal symmetry ($T$), and particle-hole symmetry ($P$). 
The classification can be represented in a table form, exhibiting Bott periodicity \cite{kitaev2009periodic}.

In the periodic table, each class is attributed its own Cartan label, which describes the symmetric space of the time evolution operator. Alternatively, the classes can be classified based 
on the sigma model target manifold or/and the symmetric space of the flattened Hamiltonian~\cite{ryu2010topological}.
For a  given symmetry class and dimensionality the system may have  $\mathbb{Z}$ or $\mathbb{Z}_2$ classification.  

 As we already discussed  in \cite{Polina2022oneD} that focused  on one-dimensional systems,  
 some symmetry classes, thought topologically distinct in general dimension, can be actually (topologically) identical for a specific dimension. 
 In the current paper, we show that similar phenomena occur in 3D systems, and focus on the peculiarities of this dimension.  
In particular,  we demonstrate that the chiral symmetric models with $\mathbb{Z}$ index can be smoothly deformed from one into another without the gap closing and the change of the topological index.   
This result is in agreement with the general table,  as obviously any chiral class can be considered as a subclass of AIII (that has chiral symmetry only).   However,  it is not immediately obvious that topological sectors with the same winding number of models with different symmetries can be smoothly connected.  
The two main goals of our paper are outlined below.
\begin{enumerate}
\item {
Construction of concrete lattice realizations of 3D topological insulators.    
To date,  the most widely studied 3D topological insulators belong to the  $\mathbb{Z}_2$ class AII~\cite{fu2007topological,qi2008topological,hasan2010colloquium,qi2011topological,hsieh2008topological,hsieh2009tunable,xia2009observation}.
The remaining classes with chiral symmetry have not been as thoroughly explored.   
A few models of topological insulators and topological superconductors have been proposed for the chiral symmetric classes DIII~\cite{hor2010superconductivity,fu2010odd}, CI~\cite{schnyder2009lattice} and AIII~\cite{PhysRevB.81.045120,xie2015surface}.
In our work,  we aim to construct the chiral symmetric models of 3D topological insulators in a unified manner.  
To do that, we couple two basic models belonging to AIII class (in the case of one dimension they correspond to SSH chains).   By using the sign ambiguity of the chiral symmetry in odd dimensions,  we define two possible chiral symmetry operators of the combined system,  that correspond to two distinct ways of coupling the AIII models.  By imposing additional symmetries on the coupling terms,  we construct the classes that have $T$ and $P$ symmetries.  The winding number in the weakly coupled case is either a sum or a difference of the winding numbers that characterize the individual blocks.  
}
\item{
Our second goal is to identify the symmetry classes that are topologically equivalent in 3D,
though this is not conspicuous from the table.    Through analyzing the stability of the surface states via adiabatic deformation of the models,  we illustrate that the three  $\mathbb{Z}$ classes AIII, DIII, and CI  are topologically equivalent.  
This implies that it is possible to construct an adiabatic path between the models of these classes without closing the gap or altering the topological invariant.    We also show that for $\mathbb{Z}_2$ classes no such path can be constructed, and therefore these two classes are distinct.}
\end{enumerate}

The paper is organized as follows: 
In Sec.~\ref{sec:symmetry} we review the non-interacting topological classification and the main 
properties of the symmetry operators.   
In Sec.~\ref{sec:basicAII},  we construct a basic model for topological insulators in class AIII.
This model serves as the building block for constructing the 3D models in all classes with chiral symmetry.
In Sec.~\ref{sec:Allchiralclasses}, we construct the 3D models in each class with chiral symmetry by coupling two basic AIII models.   We analyze the topological properties of the constructed models by computing their topological indices and also, by studying their surface states.  
In Sec.~\ref{sec:transition}  we discuss the topological equivalence of the constructed models by studying the adiabatic transformations connecting them.  
In Sec.~\ref{sec:experiment} we discuss possible experimental realization of our models. 
Finally, in Sec.~\ref{sec:summary} we summarize the results.

\section{Topological classification of  gapped systems:generalities \label{sec:symmetry}}
To have the presentation coherent and self-contained,  we briefly summarize the established properties of non-interacting topological insulators, starting with the tenfold way of topological classification.
\subsection{Tenfold way of topological classification}
The non-interacting fermionic systems can be divided into ten symmetry classes according to Altland and Zirnbauer \cite{zirnbauer1996riemannian,altland1997nonstandard} based on symmetries $C$, $T$ and $P$.
Depending on the dimension and symmetry class,  the topologically non-trivial states can be characterized either by $\mathbb{Z}$ or $\mathbb{Z}_2$ invariant as shown in Table~\ref{tab:All-Classes}.   
The tenfold classification also exhibits a periodicity as a function of spatial dimension d, known as Bott periodicity~\cite{kitaev2009periodic,stone2010symmetries}. 
In particular,  a given class has the same classification $d$ and in $\left(d+\text{period}\right)$. 
The two classes without $T$ or $P$ symmetry (A and AIII,  known as ``complex classes'') have a period of $2$ when the dimension is changed,  while the other eight classes (known as ``real classes'') have a period of $8$.
It is also worth mentioning that one can construct topological models in higher dimensions using models in lower dimensions,  or vice versa,  following the Bott clock (see Appendix~\ref{ap:Bott}). For instance, one can construct the topological models with topological invariant $\mathbb{Z}$ following the route: BDI (in $d=1$) 
$\rightarrow$ D (in $d=2$) $\rightarrow$ DIII (in $d=3$).   We will exploit this procedure to construct the equivalent of the SSH chain in three dimensions in Section \ref{sec:basicAII}.

\begin{table}
\caption{The classification table of non-interacting topological insulators and superconductors from 1D to 3D  \cite{zirnbauer1996riemannian,altland1997nonstandard}. The ten classes of single-particle Hamiltonians are classified in terms of the presence or absence of chiral symmetry ($C$), particle-hole symmetry ($P$), and time-reversal symmetry ($T$) listed from the second column to the fourth column. The absence of symmetries is denoted by $0$. The
presence of the symmetry $C$ is denoted by $1$ and the presence of the symmetry $P$ or $T$ is denoted by either $+1$ or $-1$ depending on whether the symmetry operator squares to $+1$ or $-1$.  The first column lists the Cartan labels of the ten classes. The last three columns list how the classification depends on the spatial dimension $d$. The symbols $\mathbb{Z}$ and $\mathbb{Z}_2$ indicate whether the topological phases in a given class are characterized by the integer topological invariant $\mathbb{Z}$ or by
$\mathbb{Z}_2$ invariant.  The empty table cells denote trivial classes.
\label{tab:All-Classes}}
\vspace{0.3cm}
\bgroup
\def\arraystretch{1.3} 

\begin{tabularx}{\columnwidth}{Y|YYY|cYYY}
\hline 
\hline
Class & {$C$}  & {$P$} & $T$ & $d=$ & $1$ & $2$ &  $3$\tabularnewline
\hline 
A & $0$ & $0$ & $0$ & && $\mathbb{Z}$  & \tabularnewline
AIII & $1$ & $0$ & $0$ & &  $\mathbb{Z}$ & & $\mathbb{Z}$  \tabularnewline
\hline
AI & $0$ & $0$ & $+1$ & & &  &  \tabularnewline
BDI &$1$ & $+1$  & $+1$ & & $\mathbb{Z}$ &  &  \tabularnewline
D & $0$ & $+1$ &  $0$ & & $\mathbb{Z}_2$ & $\mathbb{Z}$   &  \tabularnewline
DIII &$1$ & $+1$  & $-1$ & & $\mathbb{Z}_2$  & $\mathbb{Z}_2$  & $\mathbb{Z}$  \tabularnewline
AII &$0$ & $0$  & $-1$ & &   &$\mathbb{Z}_2$  & $\mathbb{Z}_2$ \tabularnewline
CII &$1$ & $-1$  & $-1$ & & $2\mathbb{Z}$   & &   $\mathbb{Z}_2$  \tabularnewline
C & $0$ & $-1$  & $0$  & &    & $2\mathbb{Z}$  &   \tabularnewline
CI &$1$ & $-1$  & $+1$ & &    & &   $2\mathbb{Z}$  \tabularnewline
\hline 
\hline
\end{tabularx}
\egroup
\end{table}
\subsection{Properties of symmetry operators}
Next, we review the properties of the symmetry operators that are used to classify non-interacting topological insulators and superconductors from the Table~\ref{tab:All-Classes}.  As shown in Table~\ref{tab:All-Classes}, there are three symmetry operators: chiral symmetry $C$, time-reversal symmetry $T$, and particle-hole symmetry $P$. In the single-particle Hilbert space, $C$ is unitary while $T$ and $P$ are anti-unitary~\cite{Ludwig_2015}.
The chiral symmetry operator $C$ anti-commutes with the Hamiltonian:
\begin{equation}
C H\left(\boldsymbol{k}\right) C^{-1}=-H\left(\boldsymbol{k}\right).
\end{equation}
The chiral symmetry operator is defined up to a phase, namely $e^{i\phi}C$ 
also acts as a chiral symmetry operator.

Without loss of generality,  one can restrict the phase by demanding $C^2=1$. Then according to this definition,  the operator $C$ is defined up to a sign.  The chiral symmetry is equivalent to a sub-lattice symmetry.  In order to show that one can define the projector operators onto A and B sublattices as:
\begin{align}
 \label{ChiralSublattice}
P_A = \frac{1+C}{2}, \hspace{0.5cm} 
P_B = \frac{1-C}{2} 
\end{align} 
The projectors satisfy $P_A+P_B=1$ and $P_AP_B=0$.   
In terms of those projectors, the Hamiltonian with chiral symmetry satisfies the following property: 
\begin{align}
\label{PH_commute}
P_A H P_A = P_B H P_B=0
\end{align}

This implies that there are no terms in the Hamiltonian that couple sites that belong to the same sublattice.   Note that Eq.(\ref{ChiralSublattice}) defines sublattices in a mathematical sense, which can be different from the bare labeling of the atoms in the original lattice.   It follows from  Eq.~(\ref{ChiralSublattice}) that the sign change of the chiral symmetry operator $C\rightarrow-C$ is equivalent to the swap of the labelings of the sub-lattices $P_A \leftrightarrow P_B$. According to the classification table~\ref{tab:All-Classes},  all $\mathbb{Z}$ topological systems in odd dimension have chiral symmetry.  

It is also worth mentioning that a model cannot have two chiral symmetry operators,  because their product yields a unitary symmetry operator which commutes with the Hamiltonian,  and thus the Hamiltonian can be further block-diagonalized.    Then the topological classification of the model would be determined by the symmetries of each of the blocks.

Symmetries $T$ and $P$ are anti-unitary in the single-particle Hilbert space,  and thus they can be
represented as $T = U_T K$ and $P = U_P K$, where $U_T$ and $U_P$ are unitary matrices and $K$ is complex conjugation.  A Hamiltonian  possesses symmetry $T$ if it satisfies:
\begin{equation}
U_T H^{*}\left(\boldsymbol{k}\right){U_T}^{-1}=H\left(-\boldsymbol{k}\right),
\end{equation} 
and if the model is particle-hole symmetric it satisfies: 
\begin{equation}
U_P H^{*}\left(\boldsymbol{k}\right){U_P}^{-1}=-H\left(-\boldsymbol{k}\right).
\end{equation} 
The operators $T$ and $P$ square either to $+1$ or $-1$.     

Depending on the type of time-reversal and particle-hole symmetries,  there are eight real classes,  that together with the class AIII, which has chiral symmetry only, and A class with no symmetries,  reproduce ten symmetry classes shown in the Classification Table~\ref{tab:All-Classes}.   Among them, there are five classes with chiral symmetry: AIII,  BDI, CI, CII, and DIII.  
Note that if the system possesses both $T$ and $P$ symmetries it also has chiral symmetry,  $C = P \cdot T$.  Thus the phases of the operators $T$ and $P$ can always be chosen such that  $C^2=1$. 

\subsection{Winding number\label{sec:winding}}
The chiral symmetric topological systems in odd dimensions can be characterized by a particular topological index,  known as the winding number.  It is defined as the index of the mapping between the Brillouin zone to the space of projectors onto the filled Bloch states \cite{ryu2010topological}. 
In 3D this  index can be calculated directly from the Hamiltonian through the expression \cite{PhysRevB.84.125132}
\begin{align}
\nu_{\text{3D}}=\frac{1}{48\pi^{2}}\int\mathrm{d}^{3}k \, \epsilon^{ijl}\text{Tr}\biggl[&CH_{}^{-1}\left(\partial_{k_{i}}H_{}\right)H_{}^{-1}\left(\partial_{k_{j}}H_{}\right)\nonumber\\
&H_{}^{-1}\left(\partial_{k_{l}}H_{}\right)\biggr],\label{3Dwinding}
\end{align}
where $\epsilon^{ijl}$ is the Levi-Civita  anti-symmetric tensor.  In this expression, it is implied that $C^2=1$. The chiral symmetry operator explicitly enter into this expression, which reflects the fact that chiral symmetry is necessary for the definition of the winding number. 

Note that the winding number is defined up to a sign,  that can be switched by the relabeling of the sublattices,  as follows from the fact that the relabeling changes the sign of the chiral symmetry operator $C$.  Note that because the choice of sublattices is purely a convention, such relabeling clearly does not affect any observable quantity,  and thus the models with winding number $\nu_{3D}$ and $-\nu_{3D}$ are physically equivalent. 
\section{The basic 3D topological model with chiral symmetry\label{sec:basicAII}}
\subsection{The model in class AIII}
We  construct a 3D topological model  (\ref{eq:basic3D_DIII}) 
by coupling  1D gapless wires to build a 2D topological insulator and then stack them to construct a 3D model. The details of this procedure are outlined in   Appendix~\ref{ap:BuildDIII}. 
The result is a model of a 3D topological insulator.   In momentum space the Hamiltonian reads:
\begin{align}
h_{\text{0}}=&\sum_{\boldsymbol{k}}\Big( \tau_{x}\otimes\left\{ \left[w+v\left(\cos k_{x}+\cos k_{y}+\cos k_{z}\right)\right]\sigma_{x}\right.\nonumber\\
&\left.+v\left(\sin k_{x}\sigma_{y}+\sin k_{y}\sigma_{z}\right)\right\} +v\sin k_{z}\tau_{y}\otimes\sigma_{0}\Big),\label{eq:basic3D_DIII}
\end{align}
where $k_{x},k_{y},k_{z}$ are short for $\boldsymbol{k}\cdot\boldsymbol{R}_{x}$,
$\boldsymbol{k}\cdot\boldsymbol{R}_{y}$, $\boldsymbol{k}\cdot\boldsymbol{R}_{z}$ with
$\boldsymbol{R}_{x,y,z}$ being the lattice vector,
$\sigma_{x,y,z}$ and $\tau_{x,y,z}$ are Pauli matrices and $\sigma_{0}$,
and $\tau_{0}$ are 2 by 2 identity matrices. The parameters $w$ and $v$ are real. 
 
The corresponding lattice
model in real space is depicted in Fig.~\ref{fig:layered3D}(a).
There are four spinless sites in a unit cell. The red and grey colors
represent two different layers which are denoted by the Pauli matrix
$\tau$ and the two atoms of the same color are denoted by the Pauli
matrix $\sigma$. Hoppings between layers of the same color are forbidden 
as their presence would violate chiral symmetry.  
We now discuss the symmetries of $h_0$ in some detail.
\subsection{Symmetries and winding number of $h_0$ \label{sec:symmetryofh0}}
The model  (\ref{eq:basic3D_DIII}) a has chiral symmetry with the operator  $C_{0}=\tau_{z}\sigma_{0}$.  One can define the projection operators onto the sublattices $P_A$ and $P_B$ according to (\ref{ChiralSublattice}).  
In the case of the model (\ref{eq:basic3D_DIII})  the sublattices describe atoms in layers of a different colors.   Thus from (\ref{PH_commute}), we may conclude that there are no hoppings between the atoms within one layer. 
The model
$h_0$ also has two additional symmetries: time-reversal symmetry $T_{0}=i\tau_{y}\sigma_{z}K$ ($T_{0}^{2}=-1$) and particle-hole symmetry $P_{0}=\tau_{x}\sigma_{z}K$ ($P_{0}^{2}=+1$).
However,  these two symmetries are not relevant for our construction and may be broken by extending the hopping parameter to complex numbers.  This will generate some additional terms that do not change the topological properties of the model if the change of the hopping amplitudes remains smaller than the gap which is controlled by the original hopping parameters.   From now on we will imply that the time-reversal and particle-hole symmetries in $h_0$ are weakly broken.   In this case, $h_0$ belongs to the topological class AIII, which has chiral symmetry only.  
Note,  that in one dimension we constructed the simplest AIII model by breaking time-reversal symmetry in a two-band model of BDI class that has $T^2_0=+1$.  The model with $T_0^2=-1$  in 1D requires 4 bands and can be constructed by coupling two minimal AIII models \cite{Polina2022oneD}.   This differs from the case of three dimensions,  as equivalent minimal $\mathbb{Z}$ model with $T^2_0=+1$  in 3D (class CI)  requires at least 8 bands,  and not 4 \cite{schnyder2008classification},  and thus cannot be used as a minimal building block.     

\begin{figure}[h]
\includegraphics[width=1\columnwidth]{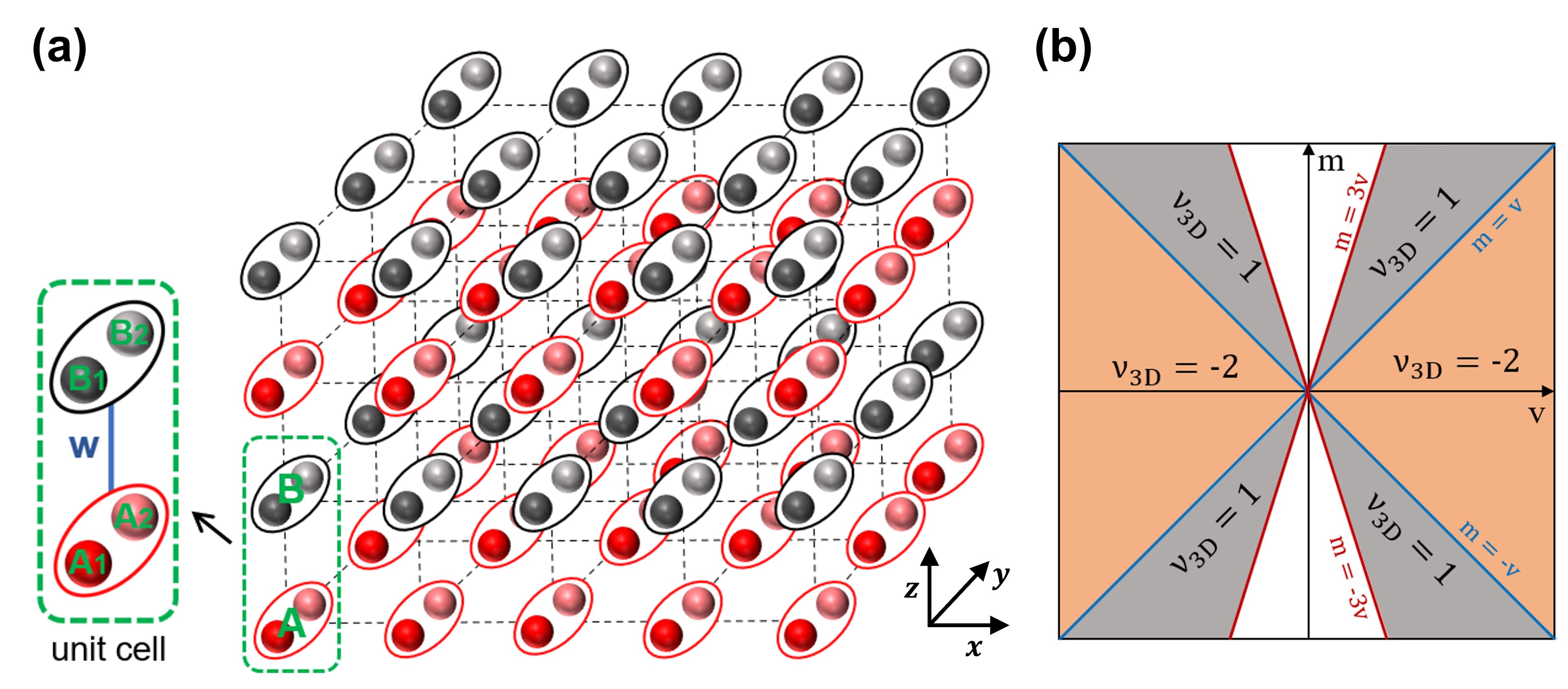}\caption{(a) The corresponding lattice model of Eq. (\ref{eq:basic3D_DIII}). There are four atoms in a unit cell. The two atoms of similar color form the space of matrices $\sigma_{0,x,y,z}$ in Eq. (\ref{eq:basic3D_DIII}). The red layer and grey layer stacked in the $z$ direction form the space of matrices $\tau_{0,x,y,z}$. 
Note that the dashed lines in (a)  are not hoppings,  as hopping within the same layer is prohibited by the chiral symmetry.  The hopping within the unit cell has the amplitude $w$, which is shown in the magnified view of the unit cell.  The intercell hoppings with the amplitude $v/2$ are illustrated in Fig.~\ref{fig:ChiralTauzHoping}(b)-(d) in the Appendix \ref{ap:BuildDIII}.
(b) The phase diagram of Eq.~(\ref{eq:basic3D_DIII}).  The blue and red lines indicate the phase boundaries. When the red line is crossed, the gap closes and reopens at one point in the Brillouin zone and when the blue line is crossed, the gap closes and reopens at three points in the Brillouin zone. The grey and orange areas indicate the topological phases with $\nu_{\text{3D}}=1$ and $\nu_{\text{3D}}=-2$, respectively.  \label{fig:layered3D}
}
\end{figure}

The topological phase diagram of Hamiltonian (\ref{eq:basic3D_DIII}) is shown in Fig.~\ref{fig:layered3D}(b).
The winding number $\nu_{\text{3D}}$ can be computed by using (\ref{3Dwinding}) and yields:

\begin{equation}
\nu_{\text{3D}}=\begin{cases}
0, & \left|w\right|>3\left|v\right|\\
1, & \left|v\right|<\left|w\right|<3\left|v\right|\\
-2, & \left|w\right|<\left|v\right|
\end{cases}\label{windingh0}
\end{equation}
According to the general argument above (see also \cite{Polina2022oneD})
the winding number we compute has a sign ambiguity.  By switching $A_1$ with $B_1$ and $A_2$ with $B_2$, the winding number  changes sign $\nu_{\text{3D}}\rightarrow -\nu_{\text{3D}}$.

\subsection{Surface states of $h_0$ \label{sec:h0surface}}
In accordance with general principles, a three dimensional topological insulator  must be equipped with a robust two-dimensional surface state. 
 We check this explicitly for our model and show that this is indeed the case, 
and  $h_0$ in its topological phase has robust surface states.  
To do it, we assume the model has a surface perpendicular to the $z$ direction and remains periodic in the other two directions.  Analogous considerations would give the same results for surfaces in other directions.

In the topological phase with $\nu_{\text{3D}}=1$,  that corresponds to $ \left|v\right|< \left|w\right|<3\left|v\right|$,  the surface states form a Dirac cone at $k_{x},k_{y}=0$ in
the surface Brillouin zone.   In the phase with $\nu_{\text{3D}}=-2$ there are two Dirac cones at the surface.  For simplicity let us focus on the topological state with $\nu_{\text{3D}}=1$.   
The effective Hamiltonian of the surface states near the Dirac point can be obtained by projecting the model Hamiltonian onto the two zero-energy surface states at $k_{x},k_{y}=0$ (see Appendix~\ref{ap:surface_states} for the details) 
\begin{equation}
h_{0}^{\text{sur}}=v\left(k_{x}\eta_{y}-k_{y}\eta_{x}\right),\label{eq:STofDIII}
\end{equation}
where $\eta_{x,y,z}$ are
Pauli matrices. 
Note that the surface Hamiltonian is chiral symmetric, and the corresponding chiral symmetry operator $C_0$ projected onto the surface states is $C^{\text{sur}}_0=\eta_z$.  The surface states are robust with respect to weak perturbations that preserve chiral symmetry $\eta_z$ and that do not close the bulk gap.  In particular, the allowed perturbations are proportional to $\eta_{x,y}$,  and thus they only shift the position of the Dirac points, but the spectrum of the surface states remains gapless.  

Note that in the one-dimensional case,  the topological surface states of $\mathbb{Z}$  insulators must be localized on the same  sublattice \cite{Polina2022oneD}.  This is related to the fact that in 1D the edge states are degenerate,  that allows to choose them to be eigenstates of the chiral symmetry operator.   This is not possible in 3D,  as the surface states are not degenerate except for the Dirac point.  At this point, they are localized on different sublattices (see Appendix \ref{ap:surface_states}) and thus their degeneracy is not protected.  This is consistent with the fact that the position of the Dirac point in $k-$ space can be shifted. 

\section{Microscopic models of three-dimensional chiral topological insulators\label{sec:Allchiralclasses}}
We now want to construct models in other classes with chiral symmetry.
We start by replicating the AIII model discussed above.
\subsection{Two uncoupled AIII models}
The general idea of construction is similar to the one proposed in reference \cite{Polina2022oneD} for one-dimensional topological insulators.  Though each of the AIII models separately has no time-reversal symmetry, we can choose them to be 
 time-reversal partners. Therefore in the full system, the time-reversal symmetry is restored.  We again emphasise that while philosophically this procedure requires the time-reversal symmetry in each of the two decoupled blocks to be (weakly) broken, mathematically it turns out not to matter.
 
 The decoupled Hamiltonian reads
\begin{equation}
H_{0}=\left(\begin{array}{cc}
h_{0} & 0\\
0 & \bar{h}_{0}\label{eq:UncoupledH0}
\end{array}\right),
\end{equation}
where $\bar{h}_{0}=T_0 h_{0} T_0^{-1}$.  Note that $h_0$ and $\bar{h}_0$ are also particle-hole counterparts of each other: $\bar{h}_{0}=-P_0 h_{0} P_0^{-1}$.  
The two blocks have the same winding number, which follows from the definition of the winding number (\ref{3Dwinding}) and the properties of the chiral symmetry operator described in Section \ref{sec:symmetry}.   Both blocks $\bar{h}_0$ and $h_0$ possess the chiral symmetry $C_0$ and thus the total model $H_0$ should also be chiral symmetric.  However,  due to a sign ambiguity of the chiral symmetry operator within each of the blocks (discussed in the Sec.  \ref{sec:symmetry})  there are two ways to define chiral symmetry in the full model.  The first and obvious  choice is to take the chiral symmetry operators of the two blocks with the same sign,  so the total operator reads
\begin{equation}
C_1=s_0\otimes C_0,\label{eq:C1}
\end{equation} 
where $s_0$ is a 2 by 2 identity matrix acting on the space of two blocks.  Using this chiral symmetry operator, the total winding number is the sum of the winding numbers of individual blocks, so the total index is $2\nu_{3D}$.

However,  if we choose the chiral symmetry operators of the individual blocks to have the opposite sign,   the combined chiral symmetry may be written as
\begin{equation}
C_2=s_z\otimes C_0,\label{eq:C2}
\end{equation} 
where $s_z$ is Pauli $z$ matrix.  This choice implies that the sign of the winding number of the second block is switched,  and therefore
the total winding number is zero. 
The ambiguity of the value of the winding number is related to the additional unitary symmetry $C_1\cdot C_2$ that is present due to the fact that the blocks $h_0$ and $\bar{h}_0$ are not coupled.  By adding coupling terms,  one can remove this ambiguity by choosing coupling that is compatible with only one of the chiral symmetry operators.  

\begin{figure}[h]
\includegraphics[width=1\columnwidth]{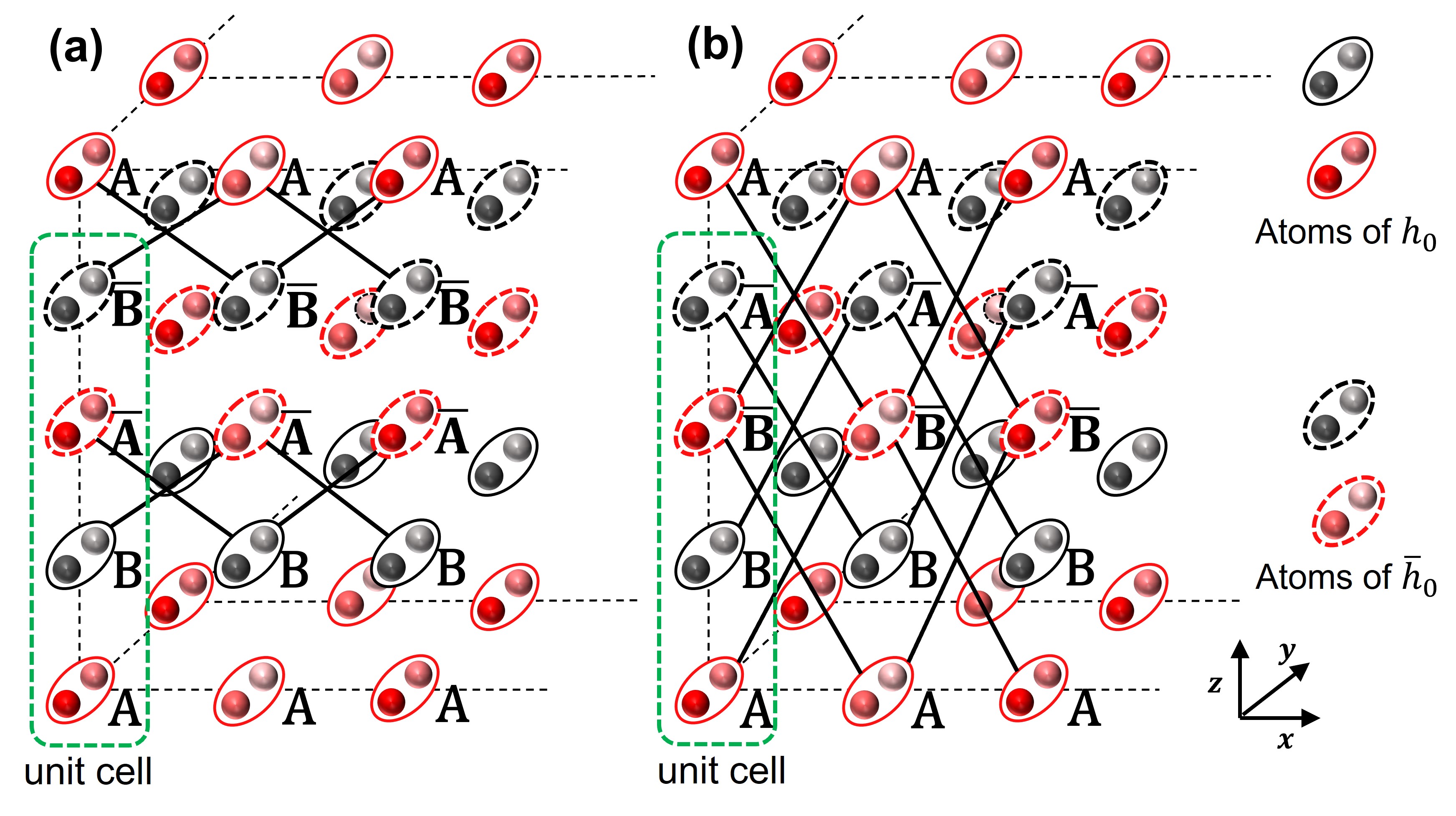}\caption{Two ways of coupling between $h_0$ and $\bar{h}_0$ in Eq.~(\ref{eq:UncoupledH0}). Atoms circled by solid lines are atoms of the model $h_0$. Atoms circled by dashed lines are atoms of the model $\bar{h}_0$.  The four atoms circled by the dashed green line form a unit cell.  The sublattices of $h_0$ are labeled by  $A$ and $B$, and the sublattices of $\bar{h}_0$ are labeled by $\bar{A}$ and $\bar{B}$.  
In (a) and (b), the sublattices of $\bar{h}_0$ have opposite labelings.
The black solid lines connecting atoms illustrate the couplings consistent with the chiral symmetry.  
The couplings in (a) preserve the chiral symmetry $C_1$ and the couplings in (b) preserve the chiral symmetry $C_2$.  Note we didn't draw all the allowed couplings here.  In (a) and (b), all couplings connecting $A$ to $\bar{B}$ and $B$ to $\bar{A}$ preserve the chiral symmetry of the coupled model.  In other words, all couplings connecting the layers of different colors (red to grey) preserve $C_1$ and all couplings connecting the layers of the same color (red to red and grey to grey) preserve $C_2$.   \label{fig:TwoCouplings}}
\end{figure}

The lattice models that correspond to the two ways of constructing the chiral symmetric system are illustrated in Fig.~\ref{fig:TwoCouplings}.  There are four atoms in the unit cell consisting of sublattices of $h_0$ labeled by $A$ and $B$  and atoms $\bar{A}$ and $\bar{B}$ that belong to the unit cell of $\bar{h}_0$.  When couplings are added the two lattices are merged into a "big lattice" with the unit cell being double that of $h_0$.   There are two ways to choose the sublattices of this ``big lattice'',  depending on the preserved chiral symmetry operator.   In particular,  for the chiral symmetry operator $C_1$ sublattices defined via the projectors (\ref{ChiralSublattice}) correspond to the choice of sublattices in the original lattice model $h_0$.  Therefore,  the coupling consistent with the operator $C_1$ connects atoms of a different color.  This is illustrated in Fig.~\ref{fig:TwoCouplings}(a).   The case of the chiral symmetry $C_2$ corresponds to the situation when the lattice labeling is switched in the second block $\bar{h}_0$.   This is illustrated in Fig.~\ref{fig:TwoCouplings}(b),  where the sublattices are labeled according to the operator $C_2$.  The allowed coupling connects atoms that have the same color in different layers.

\subsection{Time-reversal and particle-hole symmetry}
After discussing chiral symmetry,  we discuss the time-reversal symmetry and particle-hole symmetry of the uncoupled model $H_{0}$.
Since the two blocks in $H_0$ are the time-reversal counterparts of each other, $H_{0}$ is invariant with respect to two time-reversal symmetry operators that square to $\pm 1$:
\begin{equation}
\begin{cases}
T^{2}=-1: & T_{-}= s_{x} \otimes T_0= is_{x}\tau_{y}\sigma_{z}K\\
T^{2}=+1: & T_{+}= -i s_{y}\otimes T_0 = s_{y}\tau_{y}\sigma_{z}K,
\end{cases}
\label{eq:T1T2}
\end{equation}
where $T_0$ represents the time-reversal symmetry operator relating $h_0$ and $\bar{h}_0$.  The matrices $s_x$ and $s_y$  exchange the two blocks of $H_0$ and $T_0$ and transform them back. 

Similarly, one can construct two particle-hole symmetry operators:
\begin{equation}
\begin{cases}
P^{2}=+1: & P_{+}= s_x \otimes P_0 =  s_{x}\tau_{x}\sigma_{z}K\\
P^{2}=-1: & P_{-}= is_y \otimes P_0= is_{y} \tau_{x}\sigma_{z}K
\end{cases}
\label{eq:P1P2}
\end{equation}
where $P_0$ represents the particle-hole operator that relates $h_0$ and $\bar{h}_0$.  The chiral symmetry can be obtained by $C=T\cdot P$: 
\begin{equation}
\begin{cases}
 C_{1}= T_{-}P_{+}=T_{+}P_{-} =  s_{0 }\otimes C_0 \\
 C_{2}= P_{+}T_{+}= P_{-}T_{-} = s_z \otimes C_0 
\end{cases}
\label{eq:C1C2}
\end{equation}
The reason why $H_0$ has two time-reversal symmetries and two particle-hole symmetries is the presence of additional unitary symmetry of the uncoupled model.   The coupling of the two blocks in $H_0$ breaks some of the symmetries and the coupled system falls into one of the 5 classes with chiral symmetry listed in Table~\ref{tab:All-Classes} (we constrain the coupling terms to preserve one of the chiral symmetries $C_1$, $C_2$). Next, we discuss concrete examples of the coupling terms that realize models in the chiral symmetric classes from Table~\ref{tab:All-Classes}.

\subsection{The microscopic models of chiral topological
insulators}
The Hamiltonian of the coupled model can be written as:
\begin{equation}
H_{\alpha}=H_{0}+V_{\alpha},\label{eq:HplusV}
\end{equation}
where $\alpha$ denotes one of the  classes with chiral symmetry,
that are DIII, CI, AIII, CII and BDI. 
The term  $V_{\alpha}$ connects the two
blocks of $H_{0}$, preserving the symmetry
of class $\alpha$. 
The topological invariants and the symmetry operators of the corresponding model (\ref{eq:HplusV}) in each class are summarized in Table~\ref{tab:Chiral-classes}.
As mentioned before, 
the two chiral symmetries $C_1$ ($C_2$) correspond to the same (opposite) sublattice labelings of the two blocks $h_0$ and $\bar{h}_0$ in $H_0$ as illustrated in Fig.~\ref{fig:TwoCouplings}.
The presence of $C_1$ or $C_2$ only requires that the couplings must be between $A$ and $\bar{B}$ (also $B$ and $\bar{A}$).
In this case, without any further symmetries of the coupling,  the models belong to the AIII class. 
Applying symmetry constraints to the coupling allows us to construct the models of
CI, CII, DIII, and BDI classes. 
Next, we show the specific coupling terms and construct microscopic models in each class.

\begin{table}[h]
\caption{Topological classes which can be obtained by the coupled 3D models in Eq.~(\ref{eq:HplusV}). The table lists the presence or absence as well as the properties of the symmetry operators in each class. It also lists the topological index the corresponding class has in 3D. The last column shows the symmetry operators the model (\ref{eq:HplusV}) has if it belongs to the class in the same row. The time-reversal symmetry is given by (\ref{eq:T1T2}), the particle-hole by (\ref{eq:P1P2}) and the chiral symmetry is determined by $T \cdot C$ (\ref{eq:C1C2}). \label{tab:Chiral-classes}}

\vspace{0.3cm}
\def\arraystretch{1.3}
\begin{tabularx}{\columnwidth}{Y|YYY|Y|Y}
\hline 
\hline
{Class} & {$T^{2}$} & {$P^{2}$} & {$C$} & {Index} & \heading{Symmetry Operators}\tabularnewline
\hline 
DIII & $-1$ & $+1$ & $1$ & $\mathbb{Z}$ & \heading{$T_{-}$, $P_{+}$, $C_{1}$}\tabularnewline
CI & $+1$ & $-1$ & $1$ & $2\mathbb{Z}$ & \heading{$T_{+}$, $P_{-}$, $C_{1}$}\tabularnewline
AIII &  &  & $1$ & $\mathbb{Z}$ & \heading{$C_{1}$ or $C_{1}$}\tabularnewline
CII & $-1$ & $-1$ & $1$ & $\mathbb{Z}_{2}$ & \heading{$T_{-}$, $P_{-}$, $C_{2}$}\tabularnewline
BDI & $+1$ & $+1$ & $1$ & 0 & \heading{$T_{+}$, $P_{+}$, $C_{2}$}\tabularnewline
\hline 
\hline
\end{tabularx}

\end{table}

\subsubsection{Class DIII, CI}
\begin{figure}[h]
\includegraphics[width=1\columnwidth]{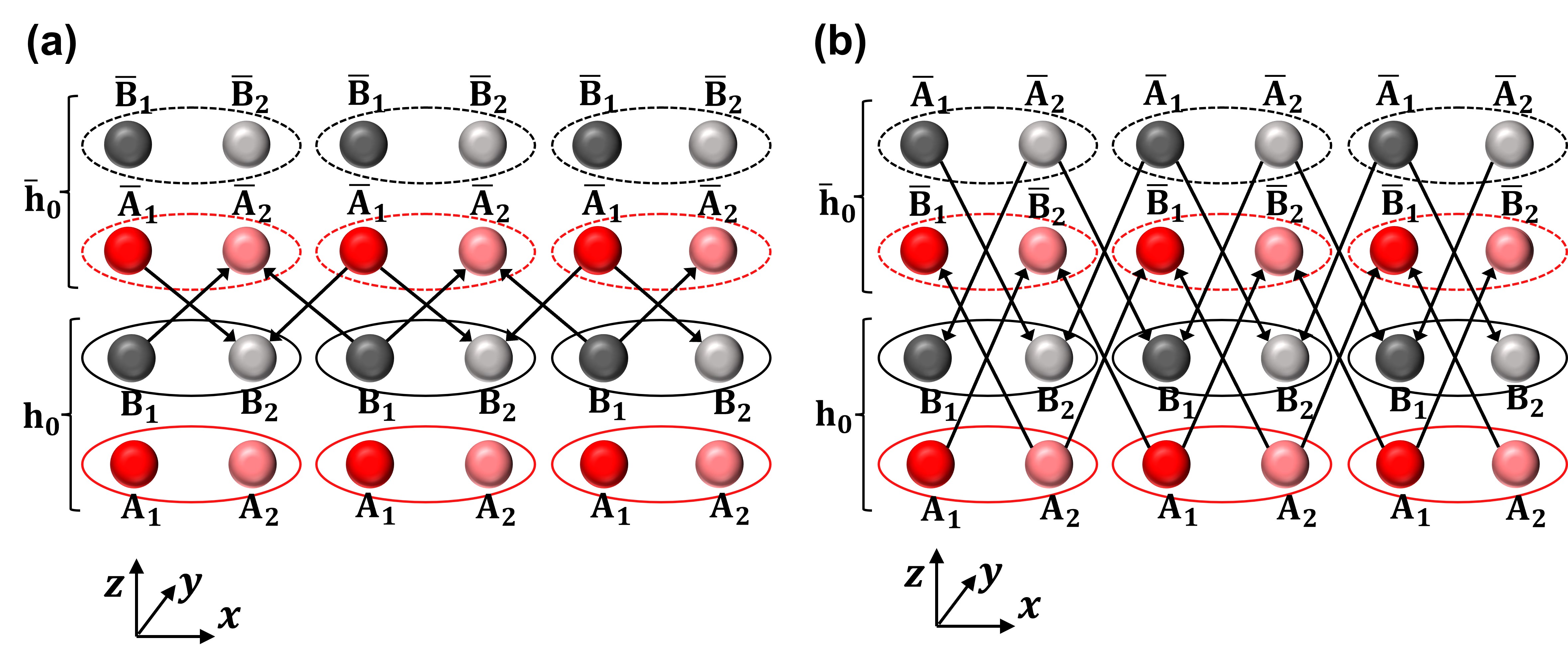}\caption{ Two ways of coupling $h_0$ and $\bar{h}_0$ in the $x$ direction. Same as the labelings in Fig.~\ref{fig:TwoCouplings}: The sublattices of $h_0$ are labeled by  $A$ and $B$. The sublattices of $\bar{h}_0$ are labled by $\bar{A}$ and $\bar{B}$. 
In (a) and (b), the sublattices of $\bar{h}_0$ have opposite labelings. The forms of couplings in (a) and (b) preserve chiral symmetry $C_1$ and $C_2$ respectively. (a) may fall into DIII class, or CI class, depending on the hopping amplitude $a$. (b) may fall into BDI class, or CII class, depending on the hopping amplitude $b$. The arrows of the hoppings indicate in which direction the hopping amplitude is $a$ and $b$. For example, in (a), if $a$ is imaginary ($a=i\left|a\right|$), the couplings are $i\left|a\right|c_{B_2,\boldsymbol{R}_n}^{\dagger}c_{A_1',\boldsymbol{R}_n}^{}+i\left|a\right|c_{B_2,\boldsymbol{R}_n}^{\dagger}c_{A_1',\boldsymbol{R}_n+\boldsymbol{R}_x}^{}+h.c.+...$} \label{fig:CouplingsInRealSpace}
\end{figure}

According to Table \ref{tab:Chiral-classes}, classes DIII and CI have chiral symmetry $C_1$,  therefore the coupling terms need to connect layers of different colors,  red to grey and vice-versa,  similar to the case in Fig.~\ref{fig:TwoCouplings}(a).
An example of the coupling structure in the classes DIII and CI is shown in Fig.~\ref{fig:CouplingsInRealSpace}(a). 
The hoppings are indicated by the arrows with amplitude $a$. 
If $a$ is real, the couplings will preserve time-reversal symmetry $T_{-}$ and particle-hole symmetry $P_{+}$. Such a model belongs to class DIII. 
In $k$ space, the couplings are written as
\begin{align}
V_{\text{DIII}}&=\frac{a}{2}\left(1+\cos k_x\right)\left(s_x\tau_x\sigma_x+s_y\tau_y\sigma_x\right)\nonumber\\
&+\frac{a}{2}\sin k_x\left(s_x\tau_x\sigma_y+s_y\tau_y\sigma_y\right),\label{eq:VDIII}
\end{align}
which is written in the basis 
\begin{equation*}
\left\{c_{A_1},c_{A_2},c_{B_1},c_{B_2},c_{\bar{A}_1},c_{\bar{A}_{2}}, c_{\bar{B}_{1}},c_{\bar{B}_2}\right\}^{T},
\end{equation*}
where we have used the short notation, $c_{A_1;k_x,k_y,k_z}\rightarrow c_{A_1}$.
The Hamilltonian $V_{\text{DIII}}$ is not the only possible model with the symmetries of DIII class -- in Table~\ref{tab:Allcouplings} we list all possible couplings in $k$ space that are compatible with the symmetries of this class. 
\begin{table}[h]
\caption{All coupling terms that are compatible with the symmetries of class DIII and BDI.  
$k$ can be any one of the momentum components $k_x$, $k_y$ and $k_z$. $f_{e}\left(k\right)$ is any real even function of $k$. $f_{o}\left(k\right)$ is any real odd function of $k$.
The coupling terms which preserve the symmetries of class CI (CII) can be obtained from the coupling terms of class DIII (BDI) by exchanging $f_{e}\left(k\right)$ and $f_{o}\left(k\right)$. } \label{tab:Allcouplings}
\vspace{0.3cm}
\def\arraystretch{1.3}
\begin{tabular*}{\columnwidth}{c|cc}
\hline 
\hline
\multicolumn{1}{c}{ {\textcolor{white}{fi}}Class {\textcolor{white}{fi}}} &  {\textcolor{white}{fi}}Functions{\textcolor{white}{fi}} & Matrices\\
\hline
\multirow{2}{*}{DIII}
&  $f_{e}\left(k\right)\times$ & $\left( s_{x}, s_{y} \right)\otimes\left(\tau_{x},\tau_{y}\right)\otimes\sigma_{x}$\\ 
\cline{2-3} 
&  $f_{o}\left(k\right)\times$ &
 {\textcolor{white}{fin}}$\left( s_{x}, s_{y} \right)\otimes\left(\tau_{x},\tau_{y}\right)\otimes\left(\sigma_{0},\sigma_{y},\sigma_{z}\right)${\textcolor{white}{find}}\\
\hline
\multirow{2}{*}{BDI}
&  $f_{e}\left(k\right)\times$ & $\left(s_{x},s_{y}\right)\otimes\left[\tau_{z}\otimes\left(\sigma_{0},\sigma_{y},\sigma_{z}\right),\tau_{0}\otimes\sigma_{x}\right]$\\ 
\cline{2-3} 
&  $f_{o}\left(k\right)\times$ 
& $\left(s_{x},s_{y}\right)\otimes\left[\tau_{0}\otimes\left(\sigma_{0},\sigma_{y},\sigma_{z}\right),\tau_{z}\otimes\sigma_{x}\right]$\\
\hline 
\hline
\end{tabular*}
\end{table}

If the hopping $a$ is imaginary: $a=i\left|a\right|$, the couplings will preserve time-reversal symmetry $T_{+}$ and particle-hole symmetry $P_{-}$, that puts the model in the class CI.  In $k$ space,  this is written as
\begin{align}
V_{\text{CI}}&=\frac{\left|a\right|}{2}\left(1-\cos k_x\right)\left(s_x\tau_x\sigma_y+s_y\tau_y\sigma_y\right)\nonumber\\
&+\frac{\left|a\right|}{2}\sin k_x\left(s_x\tau_x\sigma_x+s_y\tau_y\sigma_x\right),
\end{align}
 in the same basis as that of $V_{\text{DIII}}$ in Eq.~(\ref{eq:VDIII}).  All the allowed couplings of class CI can be obtained from the coupling terms of class DIII in Table~\ref{tab:Allcouplings} by exchanging $f_{e}\left(k\right)$ and $f_{o}\left(k\right)$.
 
 One could also take the hopping $a$ to be complex -- which would give a linear combination of $V_{\text{DIII}}$ and $V_{\text{CI}}$ in $k$ space.  Such a model still has the $C_1$ symmetry, but no longer has time reversal or particle-hole symmetries.  It would therefore be in the universality class AIII.

Suppose the coupling $V_{\text{DIII,CI}}$  is small, 
so it does not close the original energy gap determined by the parameters of $H_0$. In this case, the winding number of the weakly coupled system is the same as that of the uncoupled model $H_0$. Given the chiral symmetry is $C_1$, the total winding number will be $\nu_{\text{total}}=2\nu_{\text{3D}}$,  where $\nu_{\text{3D}}$ is the winding number of a single block $h_0$ which is given by Eq.~(\ref{windingh0}).
We will later show that the topological phases are manifested in terms of the protected surface states in Sec.~\ref{sec:Psurface},  as expected from the non-zero topological indices.

\subsubsection{Class BDI, CII}
Classes BDI and CII have chiral symmetry $C_2$, therefore the coupling terms need to connect the layers of the same color,  red to red and grey to grey similar to the case in Fig.~\ref{fig:TwoCouplings}(b).
An example of such a coupling that would be in class CII or BDI is shown in Fig.~\ref{fig:CouplingsInRealSpace}(b). 
The hoppings are indicated by the arrows with amplitude $b$. 
If $b$ is real, the couplings will preserve time-reversal symmetry $T_{+}$ and particle-hole symmetry $P_{+}$, which puts the model in the class BDI. 
In $k$ space the couplings are written as: 
\begin{align}
V_{\text{BDI}}=b\left(1+\cos k_x\right)s_x\tau_0\sigma_x+b\sin k_x s_x\tau_0\sigma_y.\label{eq:VBDI}
\end{align}
Note that $V_{\text{BDI}}$ is written in the basis 
\begin{equation*}
\left\{c_{A_1},c_{A_2},c_{B_1},c_{B_2},c_{\bar{B}_1},c_{\bar{B}_{2}},c_{\bar{A}_1},c_{\bar{A}_2}\right\}^{T},
\end{equation*} which is intended to make the form of the Hamiltonian independent on the relabelling in Fig.~\ref{fig:CouplingsInRealSpace}(b) and then the symmetry operators in Eq.~(\ref{eq:C1})-(\ref{eq:P1P2}) also apply.
Table~\ref{tab:Allcouplings} lists all the couplings which preserve the symmetries of class BDI.

If the hopping $b$ is imaginary: $b=i\left|b\right|$, the couplings will preserve time-reversal symmetry $T_{-}$ and particle-hole symmetry $P_{-}$.  This corresponds to a model in class CII. In $k$ space, the couplings are written as:
\begin{equation}
V_{\text{CII}}=\left|b\right| \left(1+\cos k_x\right)s_{y}\tau_{z}\sigma_{x}+\left|b\right| \sin k_x s_{y}\tau_{z}\sigma_{y}.
\end{equation}
$V_{\text{CII}}$ is written in the same basis as $V_{\text{BDI}}$ in Eq.~(\ref{eq:VBDI}).
All the allowed couplings of class CII can be obtained from the coupling terms of class BDI in Table~\ref{tab:Allcouplings} by exchanging $f_{e}\left(k\right)$ and $f_{o}\left(k\right)$.

If the coupling strength is small,  compared to the gap opened by the terms of the uncoupled model $H_0$,  the winding number of the coupled system is the same as in the absence of coupling.
The chiral symmetry of the models CII and BDI is $C_2$, and thus the total winding number $\nu_{\text{total}}=0$.  This is a manifestation of the fact that the classes CII and BDI do not obey $\mathbb{Z}$ classification in three dimensions. 
In particular,  the BDI class is trivial, according to Table~\ref{tab:All-Classes}, and CII is a $\mathbb{Z}_2$ topological insulator.  This can be directly proven by calculating the $\mathbb{Z}_2$ invariant of $H_{\text{CII}}$.  Later, we also verify the existence of a topologically non-trivial phase in the CII class by examining its surface states. 

Here it is also worth mentioning that by coupling the two building blocks in Eq.~(\ref{eq:UncoupledH0}), one can obtain models in both classes with non-zero winding number (DIII, CI, AIII) and classes with zero winding number (BDI, CII).  This is a result of the sign ambiguity of the topological index  as was discussed in details in Sec.~\ref{sec:winding} and in~\cite{Polina2022oneD}.  For each block,  the winding number can be either positive or negative,  thus the winding number of the coupled model can be either the sum or difference of the absolute values of the two blocks' winding numbers.  However, this logic applies only to odd dimensions.  In even dimensions the $\mathbb{Z}$ index is the Chern number,  which has no sign ambiguity.  Therefore, in even dimensions, one can obtain a model in a $\mathbb{Z}$ class by coupling a pair of models with the same Chern number.   To construct a model of $\mathbb{Z}_2$ class in even dimensions one needs to couple models with opposite Chern numbers.  Therefore one can't realize both $\mathbb{Z}$ and $\mathbb{Z}_2$ classes by coupling the same pair of models.

As before, one can also consider the case when the real-space hopping $b$ is complex, leading to a linear combination of $V_{\text{BDI}}$ and $V_{\text{CII}}$.  Such a model would have the chiral symmetry $C_2$ but no time-reversal or particle-hole symmetries and thus be in the class AIII.  Unlike the case of combing $V_{\text{DIII}}$ and $V_{\text{CI}}$ however, the winding number in this case is zero -- so although such a model would be in a class that exhibits a $\mathbb{Z}$ topological classification, it is always in the topological trivial phase.

\subsection{Protected surface states of topological insulators\label{sec:Psurface}}
Here we discuss the properties of the surface states of the constructed models in the topological phase.  
The topological nature of the models in the classes with chiral symmetry is demonstrated by the robustness of gapless surface states.
\subsubsection{Surface states of the uncoupled model}
Let us first focus on the edge states of the uncoupled model.   In Sec.~\ref{sec:h0surface} we showed that the surface states of the block $h_0$ form a Dirac cone.  Thus the model $\bar{h}_0$ also has the edge states,  as it has the same winding number as $h_0$.   If we consider a surface perpendicular to the $z$ axis,  
the effective Hamiltonian for the surface states of the uncoupled system $H_0$ (\ref{eq:UncoupledH0}) reads:
\begin{equation}
H_0^{\text{sur}}=s_{0}\otimes h_0^{\text{sur}}=v s_{0}\otimes\left(k_{x}\eta_{y}-k_{y}\eta_{x}\right).\label{eq:ST4by4}
\end{equation}
If we add coupling between the models $h_0$ and $\bar{h}_0$,  it might in general gap out the surface states.  However,  if the coupling is sufficiently small and compatible with the symmetries of a topologically non-trivial class,  the edge states will remain gapless.  We demonstrate this in the next subsections on the examples of chiral symmetric topological classes.  

\subsubsection{$C_1$ chiral topological insulators}
Here we show that the chiral symmetry $C_1$ is the key symmetry for the stability of the topological phases in $\mathbb{Z}$ classes.  In particular, given $h_0$ is in a topologically non-trivial phase,  as long as  $C_1$ is preserved,   any weak perturbation can not gap the surface states described by Eq.~(\ref{eq:ST4by4}).   In the same basis as that of Eq.~(\ref{eq:ST4by4}), the effective $C_1$ is expressed as $C^{\text{sur}}_1=s_0\eta_z$, and therefore,
the perturbation matrices which preserve this symmetry are proportional to $s_{0,x,y,z}\otimes h_{p}$. Here the matrix  $h_{p}$ preserves the chiral symmetry of a block,  $C_0=\eta_z$.  It has been shown in Sec.~\ref{sec:h0surface} that such perturbations can't gap the surface states of the block described by  Eq.~(\ref{eq:STofDIII}).   
The $s_{0,x,y,z}$ matrix couples the same or different blocks of $H_0^{\text{sur}}$ in Eq.~(\ref{eq:ST4by4}).  However,  by choosing the proper basis in the subspace where matrices $s_i$ act,  we can always block-diagonalize the perturbation and obtain:
\begin{equation}
H^{\text{sur}}=s_{0}\otimes h_0^{\text{sur}}+ (\alpha_0 s_0 +\alpha_z s_z) \otimes h_p,  
\end{equation}
Since $h_{p}$ does not open a gap in $h_{0}^{\text{sur}}$,  the perturbation $s_{0,x,y,z}\otimes h_{p}$ also leaves the edge states $H_{0}^{\text{sur}}$ gapless.
The coupling terms of the classes DIII and CII by construction preserve chiral symmetry $C_1$,  so they can't gap the surface states of the uncoupled model.  Therefore,  in the weak coupling limit,  the models with chiral symmetry $C_1$ that belong to the classes AIII, DIII, and CI are all in the topological phase characterized by protected gapless surface states.  Note,  that the other symmetries are not relevant for their protection. 

\subsubsection{$C_2$ chiral insulators\label{sec:C2chiralInsulator}}
The existence of chiral symmetry $C_2$ alone is not enough to make the whole system topological,  and additional symmetries are needed.  To prove that, consider the operator $C_2$ that in the space of the surface, states is expressed as $C^{\text{sur}}_2=s_z\eta_z$.  Consider the perturbation $s_x \otimes \eta_z$ that preserves $C_2$.  
In the basis where $s_x$ is diagonal $s_x \rightarrow s_z$,   the Hamiltonian of the surface states takes the following form:
\begin{align}
\label{eq:CII_perturbed}
H^{\text{sur}}=s_{0}\otimes h_0^{\text{sur}}+ \beta s_z \otimes \eta_z. 
\end{align}
This gaps out the surface states in each block $h^{\text{sur}}_0$ of the uncoupled model  Eq.~(\ref{eq:ST4by4}),  as this Hamiltonian describes the two massive Dirac fermions with the mass term given by $\pm \beta \eta_z$.   
That explains why the Hamiltonian of the AIII class with $C_2$ symmetry is topologically trivial in the case of weak coupling.  However, the gapless surface states may survive if the model has other symmetries.  This is the case of $\mathbb{Z}_2$ class CII,  that also has time-reversal $T_{-}$ and particle-hole $P_{-}$ symmetries.  In the space of the surface states, $T_{-}$, $P_{-}$ they are expressed as $T_{-}^{\text{sur}}=is_x\otimes \eta_y K$, $P_{-}^{\text{sur}}=is_y\otimes \eta_x K$.   With the constraint of these two additional symmetries,  no perturbations can gap the surface states.  If one further couples two copies of the CII model together, the surface Hamiltonian is expressed as $I_{4\times 4}\otimes h_0^{\text{sur}}$. Given the model is in class CII,  the symmetries $C_2^{\text{sur}}$, $T_{-}^{\text{sur}}$, $P_{-}^{\text{sur}}$ need to be preserved. In this case, there exist perturbations (e.g. $o_y\otimes s_x \otimes\eta_z$, $o_y\otimes s_y \otimes\eta_z$)  which preserve all the three symmetries, but can gap out the surface Hamiltonian.  This follows from the fact that in the basis where $o_y$ and $s_{x/y}$ are diagonal,  the effective Hamiltonian of each of CII models takes the form of Eq. (\ref{eq:CII_perturbed}) that corresponds to a gapped model.  This result shows that the CII model is characterized by topological invariant $\mathbb{Z}_2$, not $\mathbb{Z}$. 

\section{Topological equivalence of the $\mathbb{Z}$ classes \label{sec:transition}}
In this section, we discuss the topological equivalence of the chiral classes 
with the same winding number (DIII, AIII, and CI).  
Equivalence means the existence of a path along which one can smoothly transform a model from one class to another without closing the gap and changing the topological invariant.    
The equivalence of the $\mathbb{Z}$ chiral classes follows from the fact that they are characterized by the chiral symmetry operator $C_1$,  and thus the value of the winding number of a weakly coupled system does not depend on the presence of other symmetries.  In particular, if one considers a Hamiltonian of classes DIII or CI and weakly breaks time-reversal or particle-hole symmetry, the winding number would not be changed.  

This implies that one can construct a smooth path between different $\mathbb{Z}$ classes.  For instance,  one may consider the Hamiltonian $H=H_{0}+t V_{\text{DIII}}+(1-t)V_{\text{CI}} $,  where $t \in [0,1]$.  This Hamiltonian describes the interpolation between the classes CI and DIII via the AIII class.  If both coupling terms $V_{\text{DIII}}$ and $V_{\text{CI}}$ are weak,  so they do not close the gap opened by the terms in $H_0$,     a smooth transition from class DIII to class CI is achieved.  Moreover,  the winding number doesn't change along the path and if the model has a surface,  surface states won't be gapped,  as was discussed in the Sec. \ref{sec:Allchiralclasses}. 

This is consistent with the fact that all symmetry classes with chiral symmetry 
can be viewed as a special case of AIII class.  However,  because of the additional symmetries,  the existence of the path that connects the topological sector of one class
to another is not guaranteed.  In particular, 
the class CI can only have even winding numbers $2\mathbb{Z}$ due to its additional symmetries,  so class CI is only equivalent to the $2\mathbb{Z}$ sector of the classes AIII and DIII. 

It is also worth mentioning that as discussed in Sec.~\ref{sec:C2chiralInsulator}, the model $H_{\text{CII}}$ of the class CII is a $\mathbb{Z}_2$ topological insulator.  The CII model can be smoothly connected without breaking chiral symmetry to the trivial sector of the AIII model with $C_2$ chiral symmetry by adding a term that breaks the time-reversal symmetry and does not close the gap.   However,  there is no chiral symmetry preserving path between the CII insulator and  $\mathbb{Z}$ topological models with a non-zero winding number.    Along any path that connects these two models, the chiral symmetry must change from $C_2$ to $C_1$. Therefore there is a point where the chiral symmetry changes and the surface states are not protected and may be gapped. 
  
Moreover,   the models of two $\mathbb{Z}_2$ classes in 3D are distinct and cannot be adiabatically connected.  This is in full analogy with the one-dimensional case \cite{Polina2022oneD}.
Indeed,   to construct the path from the class CII to AII one needs to break the chiral symmetry but preserve the time-reversal,  which is the only symmetry of the $\mathbb{Z}_2$ class AII.  However,  such perturbations gap out the surface states of CII.   To be specific,  let us consider the Hamiltonian of the surface states  Eq.~(\ref{eq:ST4by4}) and add the perturbation that is compatible with the time-reversal symmetry $T_-$ and breaks the chiral symmetry $C_2$.   An example of such perturbation is $V= \alpha s_z\otimes\eta_z$.  This term will open a gap in the two Dirac cones described by Eq.~(\ref{eq:ST4by4}).   

\section{Proposal for a realization in a cold-atom experiment\label{sec:experiment}}
So far the discussion was purely theoretical. To study the transition between various classes 
and observe the emergence of the corresponding surface one needs to realize the constructed models in experiments. It is rather hard to control the parameters of the Hamiltonians
 in solid-state systems but seems feasible in cold atomic settings. 
 In this section, we discuss the possibility of the realization of these models in cold atom experiments.

As an illustration, consider the simplest model of a chiral topological insulator that can be used as a building block for constructing other topological classes.   The original model (\ref{eq:basic3D_DIII}) contains terms that involve the next-nearest hopping in the $z$ direction, the amplitude of that is the same as that of nearest-neighbor terms as shown in Fig.~\ref{fig:ChiralTauzHoping},  see Appendix \ref{ap:BuildDIII}.       
It is more convenient to consider a modification of this model that is more experimentally realistic and involves only  nearest-neighbor hopping in the $z$ direction only:
\begin{align}
\label{coldat}
h=\sum_{\boldsymbol{k}}  \tau_{x}\otimes \left[w+v\left(\cos k_{x}+\cos k_{y}+\cos k_{z}\right)\right]\sigma_{0} \nonumber\\
+v\sin k_{z}\tau_{y}\otimes\sigma_{0} +H_{\text{so}},
\end{align}
where $H_{\text{so}}$ is given by
\begin{align}
\label{SO_coldat}
H_{\text{so}}= v \tau_z \otimes \left(\sin k_{x}\sigma_{y} + \sin k_{y}\sigma_{x} \right).
\end{align}
The model (\ref{coldat}) is also in DIII topological class with symmetries $C=\tau_z\otimes \sigma_z$,  $T=i \tau_0 \otimes \sigma_y K$ and $P= \tau_z \otimes \sigma_x K$. It has the same topological phase diagram as the model (\ref{eq:basic3D_DIII}).  
The possible values of the winding number are given by  (\ref{windingh0}). 

We stress that although the model (\ref{coldat}) and  the model (\ref{eq:basic3D_DIII}) discussed earlier are not identical, they are interchangeable.  One could also use (\ref{coldat}) as the building block in Eq.~(\ref{eq:UncoupledH0}).
By coupling the two such models one can construct models of all the classes with chiral symmetry, as described in the main text.
The expressions of the symmetric operators and the coupling terms will change,  but this will not affect our main results.

\begin{figure}[h]
\includegraphics[width=1\columnwidth]{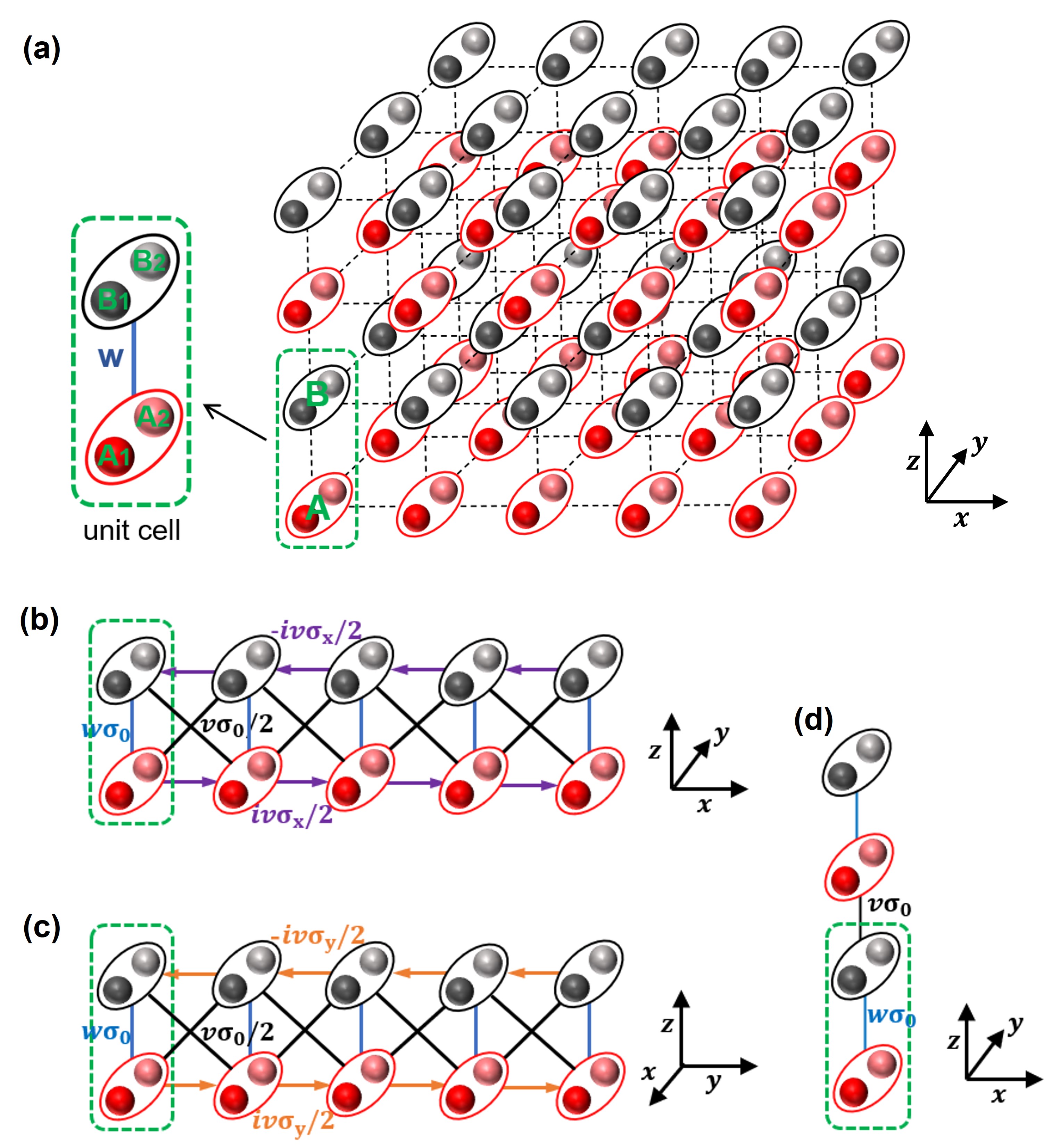}\caption{(a) is the lattice model corresponding to  Eq.  (\ref{coldat}).  The unit cell consists of four atoms.  They differ by the color that denotes the layer degrees of freedom (described by Pauli matrices $\tau_{i}$). There are two types of atoms of the same color in the unit cell, which may also describe internal degrees of freedom (described by Pauli matrices $\sigma_{i}$), e.g. spin. The dashed lines in (a)  are not hoppings. The figures in (b), (c) and (d) illustrate the hoppings in the $x$-, $y$-, and $z$-direction correspondingly.
} 
\label{fig:Coldat}
\end{figure}

The corresponding lattice model of (\ref{SO_coldat}) is illustrated in Fig.~\ref{fig:Coldat}. 
The lattice consists of atoms with internal degrees of freedom (e.g. spin) described by the Pauli matrices $\sigma_i$, and  $\tau_i$ are the Pauli matrices that describe the layers of different colors in  Fig.~\ref{fig:Coldat}.  
Besides the fact that in the $z$ direction it only has spin-independent nearest-neighbor hoppings,  another difference from the model (\ref{eq:basic3D_DIII}) is that in the model (\ref{coldat}) inter-layer hoppings are spin-independent and there are spin-dependent hopping terms (\ref{SO_coldat}) within each layer.

The model (\ref{coldat}) has spin-diagonal and spin-flipping terms (\ref{SO_coldat}) that are staggered in $z$ direction.     The spin diagonal part describes a regular hopping on a square lattice in $z,y$, and $z,x$ directions with hopping amplitude $v$ and in $z$ direction the hopping is dimerized like in the SSH model,   see Fig. ~\ref{fig:Coldat}(c).  The hopping dimerization can be obtained by superposing two counter-propagating lasers with the wavelengths $\lambda$ and $2\lambda$ \cite{Bloch2013, Leder2016} that creates a double-well optical potential for atoms.  
The realization of similar to (\ref{SO_coldat}) type of terms has been proposed theoretically \cite{Lu2020} and realized experimentally both in 2D \cite{Wu2016} and 3D \cite{Wang2021}  in the context of Weyl semimetals.  However,  these papers discuss the spin-orbit interaction with uniform hopping amplitude.   The   implementation of  the staggered part of the spin-orbit 
coupling is more challenging, and yet to be developed.

\section{Summary}
\label{sec:summary}
We constructed elementary non-interacting three-dimensional lattice models that represent all of the topological classes with chiral symmetry.  This construction enables us to study the properties of topological insulators that are not obvious 
from the general table of the topological insulators.  In addition, the models can be used as the building blocks that are needed 
to incorporate the effects of interactions or to design real materials.

 We have built the three-dimensional models from low-dimensional ones.
 The basic 3D  lattice model belongs to  AIII class.   We build it in stages. First,  by coupling a set of one-dimensional SSH chains to construct a 2D QWZ model in the D class. Then, the latter were stacked in parallel to construct an AIII model in 3D. 

 Similar to the SSH model,  the unit cell of the constructed model consists of atoms belonging to two sublattices, labeled A and B.   In addition,  there is a `psudospin' degree of freedom.   The chiral symmetry is equivalent to the sublattice symmetry,  so only hopping from different sublattices is allowed.

To construct other classes with chiral symmetry,  we couple two copies of AIII, which are time-reversal partners.  By doing so we restore the time-reversal symmetry of the full model.   The resulting model possesses all possible symmetries (chiral,  time reversal, and particle-hole).  By adding couplings that break some of the symmetries we build the realizations in all symmetry classes with chiral symmetry.

 There are two possible ways of coupling that correspond to different choices of chiral symmetry operator.  This is related to the fact that the sign of the chiral symmetry operator can be changed by the relabeling of the sublattices.    One corresponds to the case when the labels of both sublattices are the same in both models and the second choice is when in one of the models the labels are switched A $\leftrightarrow$ B.  Thus one can choose two different chiral symmetry operators in the uncoupled system.  Adding coupling removes this freedom and breaks one of the chiral symmetries.  

Moreover, choosing a specific chiral symmetry allows one to determine the winding number of a weakly coupled system.  As was proven in \cite{Polina2022oneD},  the sign of the winding number can be changed by the relabeling of the sublattices.  Therefore,  the topological index of the weakly coupled system is determined by the chiral symmetry only and can be either a sum or a difference of the winding numbers of the uncoupled building blocks,  depending on the preserved chiral symmetry operator. 

Depending on whether the couplings preserve the time-reversal and (or) particle-hole symmetries and the type of symmetry operator,  the coupled model falls into one of the five classes with chiral symmetry.  We find that the models which have the chiral symmetry $C_1$ (that corresponds to the same labeling of the two building blocks' sublattices)  fall in the classes characterized by the $\mathbb{Z}$ topological index.  Those are the classes  AIII, DIII, and CI.  

We also showed that these models (AIII, DIII, and CI)  can be adiabatically transformed one into another without closing the gap and without changing the topological index $\mathbb{Z}$.  This means that these three classes are topologically equivalent in 3D.  We illustrated this equivalence by studying the surface states in topological phases and by proving that they remain gapless and robust as long as the chiral symmetry is preserved and the perturbations do not close the bulk gap.  

The models that are characterized by the second chiral symmetry operator $C_2$ belong to one of the AIII, CII, and BDI  classes.  
BDI class is topologically trivial in 3D,  and the CII class is characterized by the $\mathbb{Z}_2$ topological index.   By studying the surface states of the constructed models we explicitly demonstrated that  the two  
$\mathbb{Z}_2$ classes  CII and AII are not topologically equivalent.

\section*{acknowledgments}
D. G. is supported by ISF-China 3119/19 and ISF 1355/20.
P.M.  acknowledges support from the Israel Council for Higher Education Quantum Science and Technology Scholarship.  We are grateful to P. Ostrovsky and L. Khaykovich for the valuable discussions.

\appendix

\section{Building the 3D model in class DIII in real space
\label{ap:BuildDIII}}
This section shows the construction of the 3D topological model with a nonzero winding number in class DIII in the route BDI(1D)$\rightarrow$D(2D)$\rightarrow$DIII(3D).
\subsection{Bott clock of the 10 symmetry classes\label{ap:Bott}}
\begin{figure}[h]
\includegraphics[width=0.8\columnwidth]{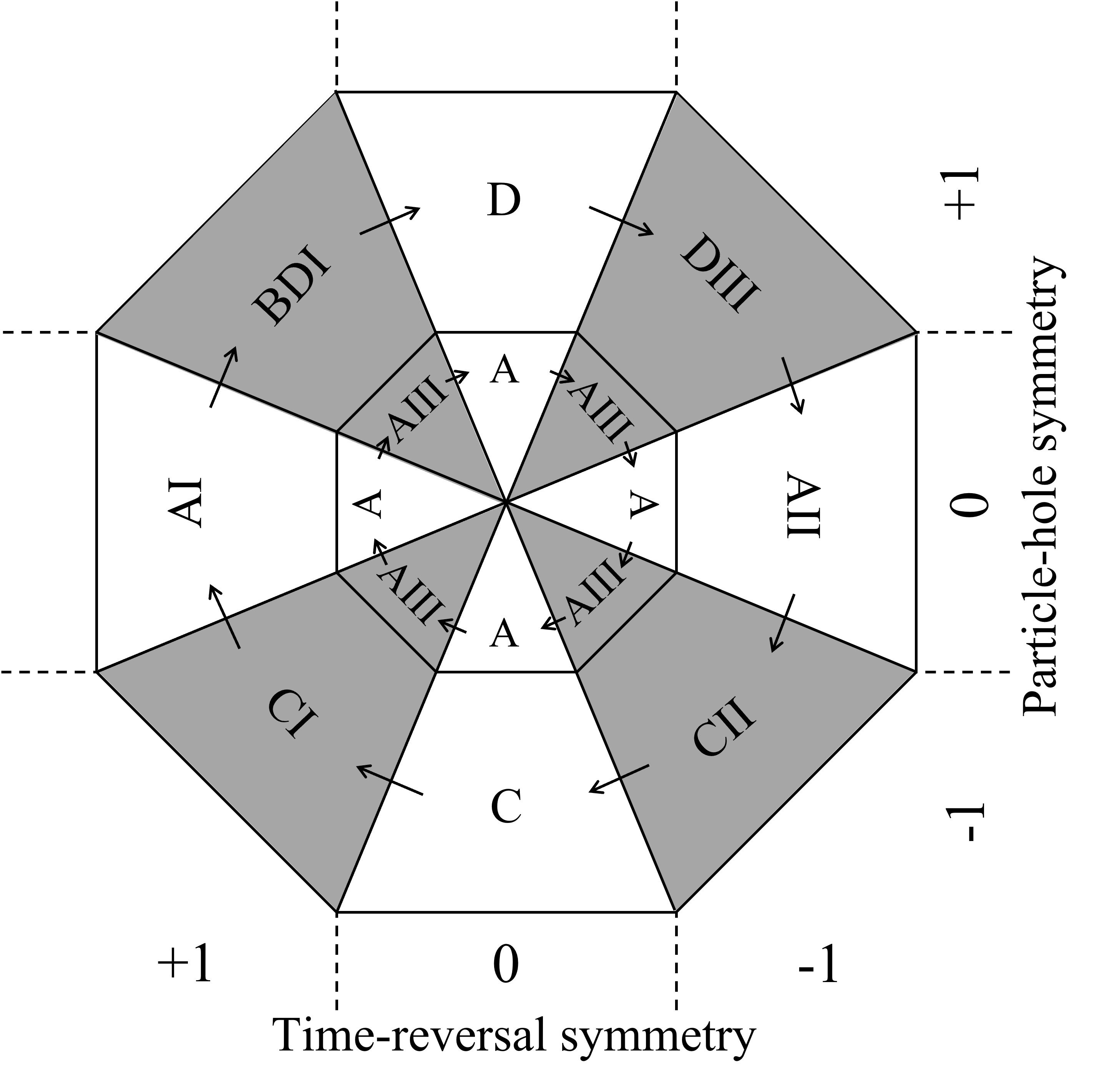}\caption{The Bott clock. 
The arrows point to the classes with the same topological classification as the previous classes when increasing the dimension $d\rightarrow d+1$. 
The grey and the white partitions represent classes with and without chiral symmetry, respectively.
$+1$ and $-1$ denote the presence of time-reversal symmetry $T$ or particle-hole symmetry $P$ and they indicate whether the symmetry operator squares to $+1$ or $-1$.
The eight real classes are in the outer circle of the clock with a period of $8$. 
They all have at least one of the $T$ or $P$ symmetry.
The two complex classes are in the inner circle and have no symmetry $T$ or $P$.
Following the circle, they are related only to each other and therefore the inner cycle has a period of $2$. 
\label{fig:BottClock}}
\end{figure}

We start with a brief review of the known results~\cite{kitaev2009periodic,stone2010symmetries}, 
which are presented here for the sake of a reader.
Fig.~\ref{fig:BottClock}(a) is the Bott clock which shows the Bott periodicity of the tenfold classification of topological insulators.
It shows that one can obtain a model of dimension $d+1$ from a model in dimension $d$ 
 without closing the gap.  Following the general idea of construction models following the Bott clock \cite{Teo2010},  we can obtain the 3D model in DIII class starting from the model of 
 BDI class (in 1d)$\rightarrow$ class D (in 2d)$\rightarrow$ class DIII (in 3d).

\subsection{From class BDI in 1d to class D in 2d}

The simplest BDI model in 1D is {represented by the SSH model~\cite{SSHmodel}.  Its Hamiltonian
in $k$ space is given by:
 
\begin{equation}
H_{\text{SSH}}\left(k_x\right)=\left(w+v\cos k_{x}\right)\sigma_{x}+v\sin k_{x}\sigma_{y},\label{eq:SSH}
\end{equation}
where $\sigma_{x,y,z}$ are Pauli matrices. The SSH model has the
time-reversal symmetry $T=K$ with $T^{2}=+1$, the particle-hole
symmetry $P=\sigma_{z}K$ with $P^{2}=+1$ and the chiral symmetry
$C=\sigma_{z}$. $H_{\text{SSH}}$ has non-zero winding number $\left|\nu_{1D}\right|=1$
when $\left|w\right|<\left|v\right|$.

When $w=-v$, the gap closes near $k_{x}=0$, and around this point the SSH chain
has two propagating modes described by a Dirac Hamiltonian: $h_{\text{SSH}}^{\text{eff}}=vk_{x}\sigma_{y}$.
We denote these two modes as left-propagating and right-propagating modes as shown in Fig. \ref{fig:CoupleSSH}(a). One can put together
such $N$ parallel SSH chains as shown in Fig. \ref{fig:CoupleSSH}(b), and introduce new couplings between chains in the $y$-direction.  The couplings only couple
the right-moving mode of the $n$-th chain and the left-moving mode of the $\left(n-1\right)$-th chain. As a result, only the two modes on the two edges remain uncoupled and gapless: the left-moving mode of the $N$-th chain and the right-moving mode of the 1st chain.   In this way, we construct the model that has a bulk gap and a gapless surface state, which is a signature of a topological phase.  
\begin{figure}[h]
\centering{}\includegraphics[width=1\columnwidth]{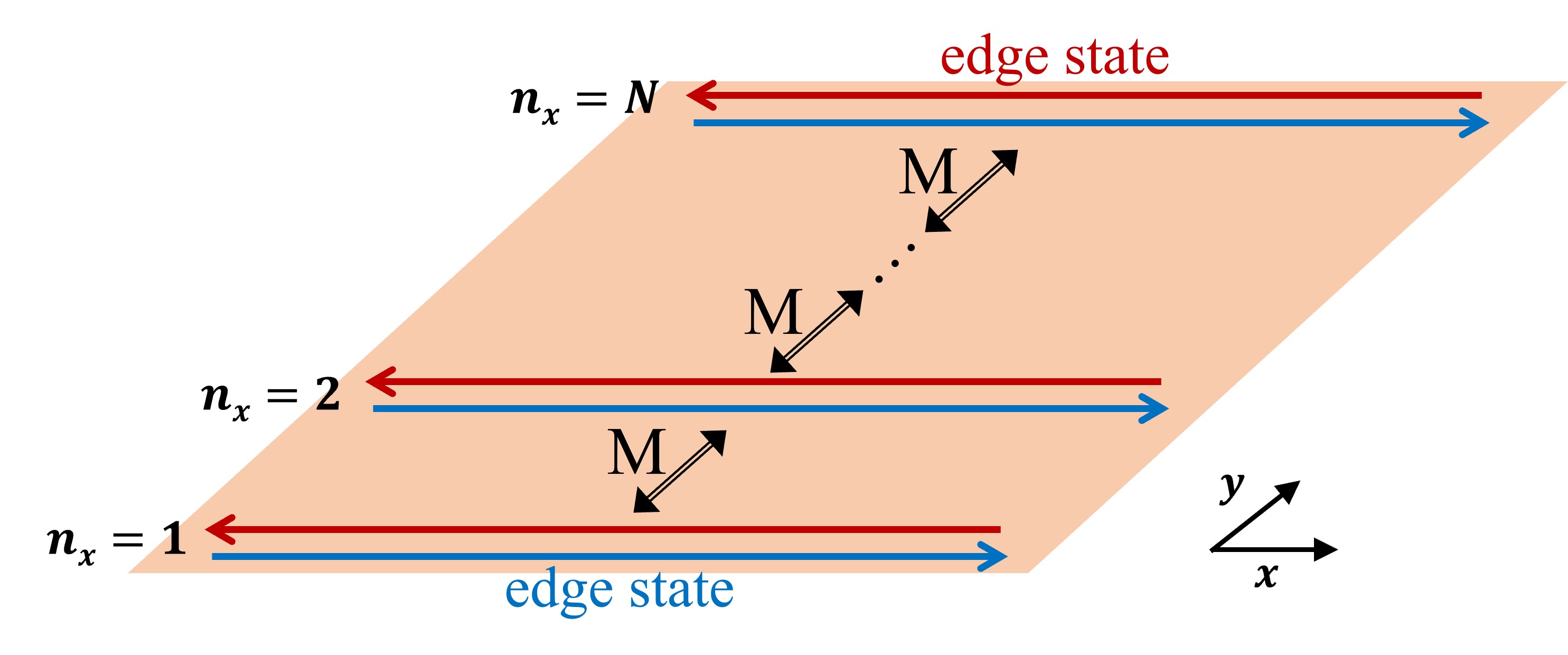}\caption{(a) The two conducting modes of a single SSH chain. (b) The 2-dimensional
model in class D by coupling $N$ SSH chains.\label{fig:CoupleSSH}}
\end{figure}
The couplings between chains that gap out the right- and left-moving modes can be written as: $M=\frac{\alpha}{2}\left(\sigma_{x}-i\sigma_{z}\right)\left|n_{y}\right\rangle \left\langle n_{y}+1\right|$.
Now we generalize the constructed model and consider the case when $w \neq v$.  We add the coupling $M$ to (\ref{eq:SSH}) and write the full  2-dimensional
model in $k$ space:
\begin{align}
H_{D}\left(k_{x},k_{y}\right)=&\left[w+v\left(\cos k_{x}+\cos k_{y}\right)\right]\sigma_{x}\nonumber\\
&+v\sin k_{x}\sigma_{y}+v\sin k_{y}\sigma_{z},\label{eq:HclassD}
\end{align}
where we take $\alpha=v$ to avoid too many parameters and simplify our model.  The term $v\sin k_{y}\sigma_{z}$
breaks the time-reversal symmetry and the chiral symmetry of Eq. (\ref{eq:SSH}).
Thus the model (\ref{eq:HclassD}) has the particle-hole $P=\sigma_{z}K$ symmetry only and thus belongs to the class D, characterized by the Chern number $Q_{2D}$ \cite{ryu2010topological}.
It can be computed explicitly and yields $\left|Q_{2D}\right|=1$ when $\left|w\right|<2\left|v\right|$,  so the model that we constructed is indeed topologically non-trivial.  This 2D model is the QWZ model~\cite{QWZmodel}.

\subsection{From 2D class D to 3D class DIII}
Next, we use the two-dimensional model Eq. (\ref{eq:HclassD})
to construct the three-dimensional model in class DIII, following the same logic as that in the last section. When $w=-2v$, the gap closes
near $k_{x}=0,k_{y}=0$. In the vicinity of this point Eq. (\ref{eq:HclassD}) has two conducting
modes in the bulk which are the eigenstates of $h_{\text{D}}^{\text{eff}}=v\left(k_{x}\sigma_{y}+k_{y}\sigma_{z}\right)$
The eigenstates of $h_{\text{D}}^{\text{eff}}=v\left(k_{x}\sigma_{y}+k_{y}\sigma_{z}\right)=v\left|k\right|\left(\cos\theta\sigma_{y}+\sin\theta\sigma_{z}\right)$
are 

\begin{equation} 
+v\left|k\right|:\psi_{a}=\frac{1}{2}\left(\begin{array}{cc}
1-ie^{i\theta}, & 1+ie^{i\theta}\end{array}\right)^{T}
\end{equation}

\begin{equation}
-v\left|k\right|:\psi_{b}=\frac{1}{2}\left(\begin{array}{cc}
1+ie^{i\theta}, & 1-ie^{i\theta}\end{array}\right)^{T}
\end{equation}

They are the eigenstates of the particle-hole symmetry operator: $P=\sigma_{z}K$, $P\psi_{a}=\alpha\psi_{a}$, $P\psi_{b}=\alpha\psi_{b}$,
where $\alpha$ is a complex number.

To build a DIII model, $2$ by $2$ matrices are not sufficient,
since it is not possible to construct a Hamiltonian that would preserve both time-reversal and particle-hole symmetries of this class in terms of Pauli matrices. Therefore, we need to duplicate the 2D lattice
Eq. (\ref{eq:HclassD}):

\begin{align}
\label{eq:ModelDCoup}
\tilde{H}_{D}=\tau_{z}\otimes H_{D}=&\tau_{z}\otimes\left(\left[w+v\left(\cos k_{x}+\cos k_{y}\right)\right]\sigma_{x}\right.\nonumber\\
&\left.+v\sin k_{x}\sigma_{y}+v\sin k_{y}\sigma_{z}\right),  
\end{align}
This can be obtained by putting two 2D models (\ref{eq:HclassD}) with parameters $w,v$
and $-w,-v$ together.  Then we can have new symmetry operators. Together
with the original particle-hole operator, they are
\begin{align}
T&=i\tau_{y}\sigma_{z}K,\ T^{2}=-1;.\nonumber\\
P&=\sigma_{z}K,\ P^{2}=+1;\nonumber\\
C&=\tau_{y}.\label{eq:SymmetryOf2dD}
\end{align}
This group of operators determines the class DIII.  The model (\ref{eq:ModelDCoup}) also has additional symmetries.  For instance, it is invariant with respect to $C=\tau_{x}$, $T=\tau_{x}\sigma_{z}K$. 
but these symmetries will be broken by introducing the new coupling terms.  When $w=-2v$, near the gap closing point $k_{x}=0,k_{y}=0$, there are
now four conducting modes.
There are two degenerate states that have the energy $+v\left|k\right|$: 
$\left\{ \psi_{a},0\right\} ^{T},\left\{ 0,\psi_{b}\right\}^{T}$  
and two degenerate states that have  the energy $-v\left|k\right|$: 
$\left\{ \psi_{b},0\right\} ^{T},\left\{ 0,\psi_{a}\right\} ^{T}$. 
Now for simplicity, we fix $k_x=0$ and write the four eigenstates as $+v k_y$: $\psi_A=\left\{-i,0,0,1\right\}^T/\sqrt{2}$, $\psi_B=\left\{i,0,0,1\right\}^T/\sqrt{2}$ and $-v k_y$: $\psi_C=\left\{0,-i,1,0\right\}^T/\sqrt{2}$, $\psi_D=\left\{0, i,1,0\right\}^T/\sqrt{2}$  (a unitary transformation has been acted on the two-states degenerate space).
$\psi_{A}$ and $\psi_{C}$ can form a space which the symmetries
in Eq. (\ref{eq:SymmetryOf2dD}) apply, so can $\psi_{B}$ and $\psi_{D}$, as illustrated in Fig. \ref{fig:Couple2dD}(a). 
Next, we stack $N$ layers of 2D $\tilde{H}_{D}$ on top of each
other in the $z$ direction. And we introduce the couplings between layers
which gap all the states in the bulk, leaving only the state $\psi_{A},\psi_{C}$
on the top and e $\psi_{B},\psi_{D}$ on the bottom. The requirement can be satisfied by the coupling
matrix $M=\frac{\alpha}{2}\left(\tau_{z}\sigma_{x}-i\tau_{x}\sigma_{0}\right)\left|n_{z}\right\rangle \left\langle n_{z}+1\right|$.

\begin{figure}[h]

\centering{}\includegraphics[width=1\columnwidth]{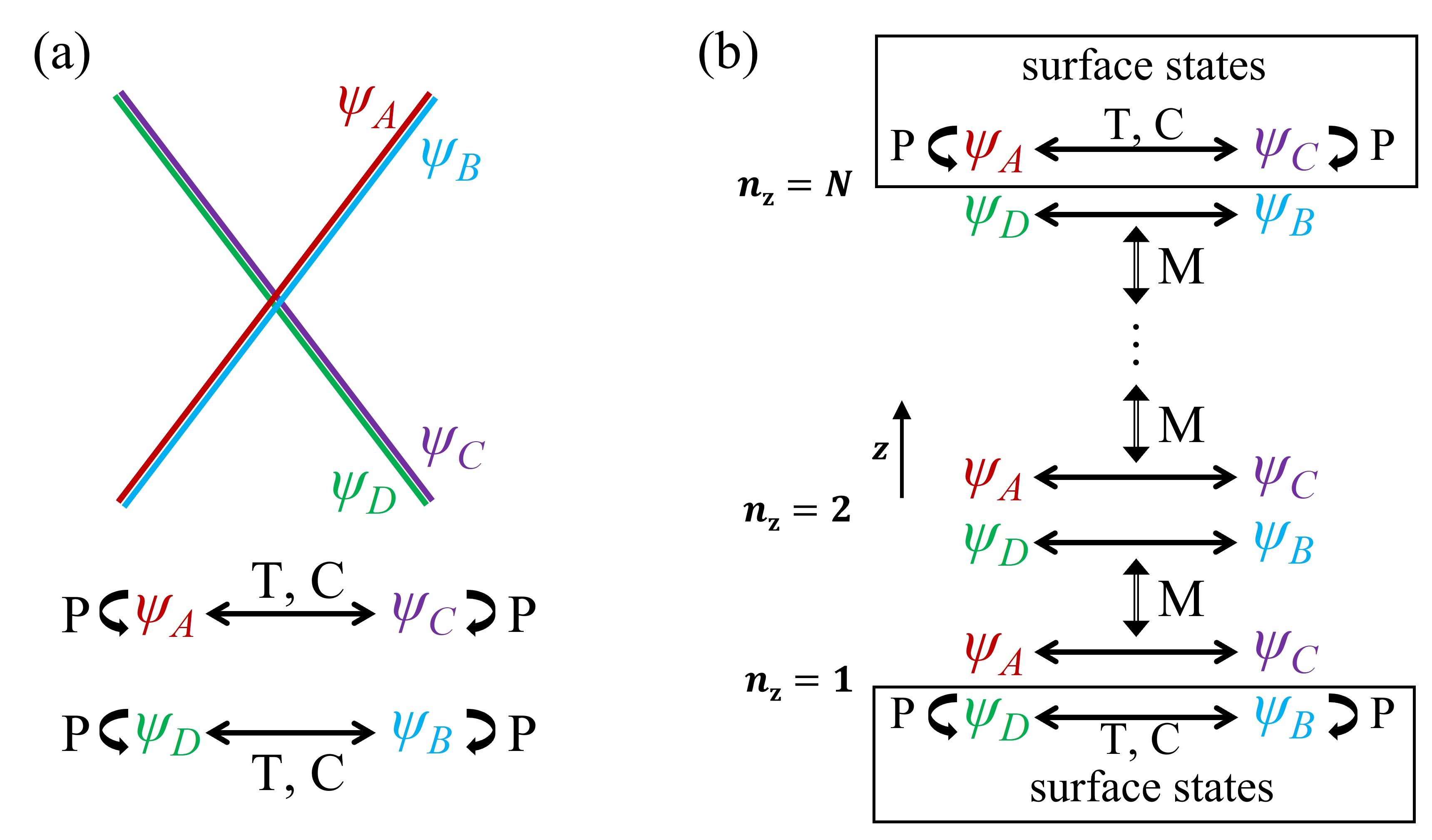}\caption{(a) An illustration of bulk states of $\tilde{H}_{D}$ near $k_{x}=0,k_{y}=0$.
Note that $\psi_{A,B,C,D}$ depends on $k_{x},$$k_{y}$, here the
horizonal is one direction in the $k_{x}-k_{y}$ plane. $\psi_{A}$($\psi_{C}$)
and $\psi_{B}$($\psi_{D}$) are completely degenerate. $\psi_{A,B,C,D}$
are all self-particle-hole symmetric. $\psi_{A}$($\psi_{B}$) and
$\psi_{C}$($\psi_{D}$) are chiral and time-reversal symmetric of
each other. (b) The 3-dimensional model in class DIII by stacking
$N$ class D layers. \label{fig:Couple2dD}}
\end{figure}

Next, we generalize the model to the case when  $w \neq - 2v$.  So we add the coupling terms to the 2D model (\ref{eq:ModelDCoup}).  
As a result, we obtain a model which is bulk-gaped but has gapless
surface states. It also preserves the symmetry of class DIII listed
in Eq. (\ref{eq:SymmetryOf2dD}). We write the total Hamiltonian of
this model in $k$ space:
\begin{align}
&H_{\text{DIII}}\left(k_{x},k_{y},k_{z}\right)=\nonumber\\
&\tau_{z}\otimes\left(\left[w+v\left(\cos k_{x}+\cos k_{y}+\cos k_{z}\right)\right]\sigma_{x} \right.\nonumber\\
&\left. +v\sin k_{x}\sigma_{y}+v\sin k_{y}\sigma_{z}\right)+v\sin k_{z}\tau_{x}\otimes\sigma_{0},\label{eq:H3dDIII01}
\end{align}
where we also take $\alpha=v$ to avoid too many parameters.  
After the rotation $\tau_z \rightarrow \tau_x , \tau_x \rightarrow  \tau_y$
Eq.\ref{eq:H3dDIII01} becames Eq.\ref{eq:basic3D_DIII} of the main text.

The model $H_{\text{DIII}}$
is topological when $\left|w\right|<3\left|v\right|$, its winding
number is 
\begin{equation}
\nu_{\text{3D}}=\begin{cases}
0, & \left|w\right|>3\left|v\right|\\
1, & \left|v\right|<\left|w\right|<3\left|v\right|\\
-2, & \left|w\right|<\left|v\right|
\end{cases}
\end{equation}

The corresponding lattice model in real space is
\begin{align}
H_{\text{DIII}}=H_0+H_x+H_y+H_z,\label{eq:FirstDIII}
\end{align}
with
\begin{align}
H_0= w\sum_{n} \left(c_{A,n}^{\dagger}c_{B,n}-c_{C,n}^{\dagger}c_{D,n}\right)+h.c.,\label{eq:FirstDIIIH0}
\end{align}
\begin{align}
H_x= v\sum_{n} \left(
c_{A,n+x}^{\dagger}c_{B,n}+c_{C,n+x}^{\dagger}c_{D,n}\right)+h.c.,\label{eq:FirstDIIIHx}
\end{align}
\begin{align}
H_y= \frac{v}{2}&\sum_{n} \left[
c_{A,n+y}^{\dagger}c_{B,n}^{}
+c_{B,n+y}^{\dagger}c_{A,n}^{}\right.\nonumber\\
&-\left(c_{C,n+y}^{\dagger}c_{D,n}^{}
+c_{D,n+y}^{\dagger}c_{C,n}^{}\right)\nonumber\\
&-i\left(c_{A,n+y}^{\dagger}c_{A,n}^{}
+c_{B,n+y}^{\dagger}c_{B,n}^{}\right)\nonumber\\
&\left.+i\left(c_{C,n+y}^{\dagger}c_{C,n}^{}
+c_{D,n+y}^{\dagger}c_{D,n}^{}\right)\right]
+h.c.,\label{eq:FirstDIIIHx}
\end{align}
\begin{align}
H_z= \frac{v}{2}&\sum_{n} \left[
c_{A,n+z}^{\dagger}c_{B,n}^{}
+c_{B,n+z}^{\dagger}c_{A,n}^{}\right.\nonumber\\
&-\left(c_{C,n+z}^{\dagger}c_{D,n}^{}
+c_{D,n+z}^{\dagger}c_{C,n}^{}\right)\nonumber\\
&-i\left(c_{A,n+z}^{\dagger}c_{C,n}^{}
+c_{C,n+z}^{\dagger}c_{A,n}^{}\right)\nonumber\\
&\left.-i\left(c_{B,n+z}^{\dagger}c_{D,n}^{}
+c_{D,n+z}^{\dagger}c_{B,n}^{}\right)\right]
+h.c.,\label{eq:FirstDIIIHx}
\end{align}
where $n,x,y,z$ are short for $\boldsymbol{R}_n,\boldsymbol{R}_x,\boldsymbol{R}_y,\boldsymbol{R}_z$. $\boldsymbol{R}_n$ is the position of the unit cell, and
$\boldsymbol{R}_x$,$\boldsymbol{R}_y$,$\boldsymbol{R}_z$ are the primitive
vectors connecting the unit cells. The lattice is shown in Fig. \ref{fig:LatticeModel}(a).
There are four atoms in one unit cell. The whole lattice is composed of SSH chains in $x$ direction. The two atoms of similar color (for example, A and B) form the space of matrices $\sigma_{0,x,y,z}$ in Eq. (\ref{eq:H3dDIII01}). The red layer and grey layer stacked in the $z$ direction form the space of matrices $\tau_{0,x,y,z}$. The black lines connecting the atoms in Fig. \ref{fig:LatticeModel}(a) are not the hoppings. The hoppings in Eq. (\ref{eq:H3dDIII01}) are drawn in Fig. \ref{fig:LatticeModel}(b) and (c). Fig. \ref{fig:LatticeModel}(b)
shows the hoppings of the red layer in $x-y$ plane and Fig. \ref{fig:LatticeModel}(c) shows the hoppings in $x-z$ plane where one can find all the hoppings
between layers in the $z$-direction.

\begin{figure}[h]

\centering{}\includegraphics[width=1\columnwidth]{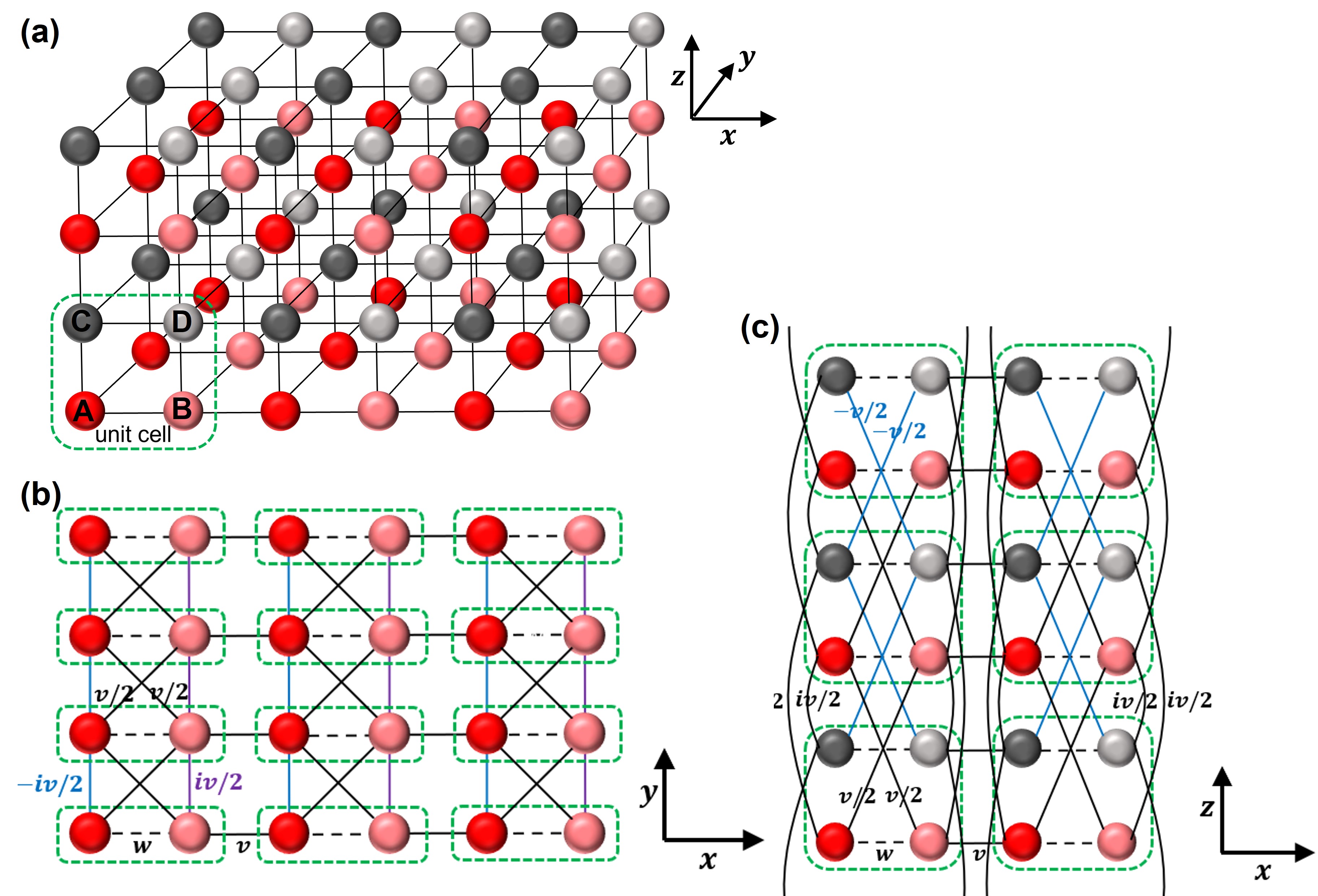}\caption{(a) is the corresponding lattice model of Eq. (\ref{eq:H3dDIII01}).
There are four atoms in one unit cell. The two atoms of similar color are described by the space of matrices $\sigma_{0,x,y,z}$ in Eq. (\ref{eq:H3dDIII01}).
The red layer and grey layer stacked in the $z$ direction form the space of matrices $\tau_{0,x,y,z}$. (b) and (c) show all the hoppings.
(b) shows the hoppings of the red layer in the $x-y$ plane. The hopping terms within the grey layer are the same as in the red layer but with $w\rightarrow-w$,
$v\rightarrow-v$. (c) shows the hoppings in the $x-z$ plane where you can find all the hoppings between layers in the $z$-direction. \label{fig:LatticeModel}}
\end{figure}
For convenience in expressing the chiral symmetry, we can
do a unitary transformation to $H_{\text{DIII}}$ in Eq.~(\ref{eq:H3dDIII01}) and transform it to
\begin{align}
&h_{0}\left(k_{x},k_{y},k_{z}\right)=\nonumber\\
& \tau_{x}\otimes\left\{ \left[w+v\left(\cos k_{x}+\cos k_{y}+\cos k_{z}\right)\right]\sigma_{x}\right.\nonumber\\
&\left.+v\left(\sin k_{x}\sigma_{y}+\sin k_{y}\sigma_{z}\right)\right\} +v\sin k_{z}\tau_{y}\otimes\sigma_{0}.\label{eq:H3dDIII02}
\end{align}
After the unitary transformation, the three symmetry operators become
\begin{align}
T_0&=i\tau_{y}\sigma_{z}K,\ T_0^{2}=-1;\nonumber\\
P_0&=\tau_{x}\sigma_{z}K,\ P_0^{2}=+1;\nonumber\\
C_0&=\tau_{z}.\label{eq:SymmetryOf2dD-1}
\end{align}
Here we can still take $\tau_{z}$ as the Pauli matrix that represents
two layers. Then we have a new lattice of which the chiral symmetry
is the sublayer symmetry. The corresponding lattice model is given in Fig.~\ref{fig:ChiralTauzHoping}.
The Hamiltonian in real space can be written as
\begin{align}
H_{\text{DIII}}=H_0+H_x+H_y+H_z,\label{eq:H3dDIIIreal}
\end{align}
with
\begin{align}
H_0=w\sum_{n}\sigma_x c_{A,n}^{\dagger}c_{B,n}^{}+h.c.,\label{eq:H3dDIIIH0}
\end{align}
\begin{align}
H_x&=\frac{v_x}{2}\sum_{n}\left[\sigma_x \left(c_{A,n}^{\dagger}c_{B,n+x}^{}+c_{A,n}^{\dagger}c_{B,n-x}\right)\right.\nonumber\\
&\left.-i\sigma_y \left(c_{A,n}^{\dagger}c_{B,n+x}-c_{A,n}^{\dagger}c_{B,n-x}\right)\right]+h.c.,\label{eq:H3dDIIIHx}
\end{align}
\begin{align}
H_y&=\frac{v_y}{2}\sum_{n}\left[\sigma_x \left(c_{A,n}^{\dagger}c_{B,n+y}+c_{A,n}^{\dagger}c_{B,n-y}\right)\right.\nonumber\\
&\left.-i\sigma_z \left(c_{A,n}^{\dagger}c_{B,n+y}-c_{A,n}^{\dagger}c_{B,n-y}\right)\right]+h.c.,\label{eq:H3dDIIIHy}
\end{align}
\begin{align}
H_z&=\frac{v_z}{2}\sum_{n}\left[\sigma_x \left(c_{A,n}^{\dagger}c_{B,n+z}+c_{A,n}^{\dagger}c_{B,n-z}\right)\right.\nonumber\\
&\left.-\sigma_0 \left(c_{A,n}^{\dagger}c_{B,n+z}-c_{A,n}^{\dagger}c_{B,n-z}\right)\right]+h.c.,\label{eq:H3dDIIIHz}
\end{align}
The whole lattice model is drawn in Fig.~\ref{fig:ChiralTauzHoping}(a), in which the zoom-in figure shows the unit cell.
In each unit cell, there are two groups of atoms labeled  A and B. The A-B sublattices form the space of  $\tau_{0,x,y,z}$. The hoppings in Eq.~(\ref{eq:H3dDIIIH0}) are represented by the blue line connecting the A and B groups in the unit cell. The hoppings between the unit cells in the $x,y,z$ direction ($H_{x,y,z}$) are given in Fig.~\ref{fig:ChiralTauzHoping}(b),(c),(d) respectively.
The chiral symmetry requires that the hoppings exist only between A and B sublattices. In Eq.(\ref{eq:H3dDIIIHx})-(\ref{eq:H3dDIIIHz}), we have tuned the hopping amplitudes in each direction to be different.
Now the condition of the phase transition between the trivial phase and topological phase changes from $\left|w\right|=3\left|v\right|$ to $\left|w\right|=\left|v_x+v_y+v_z\right|$ (here $v_x$,$v_y$,$v_z$ are all positive or negative).

\begin{figure}[h]

\centering{}\includegraphics[width=1\columnwidth]{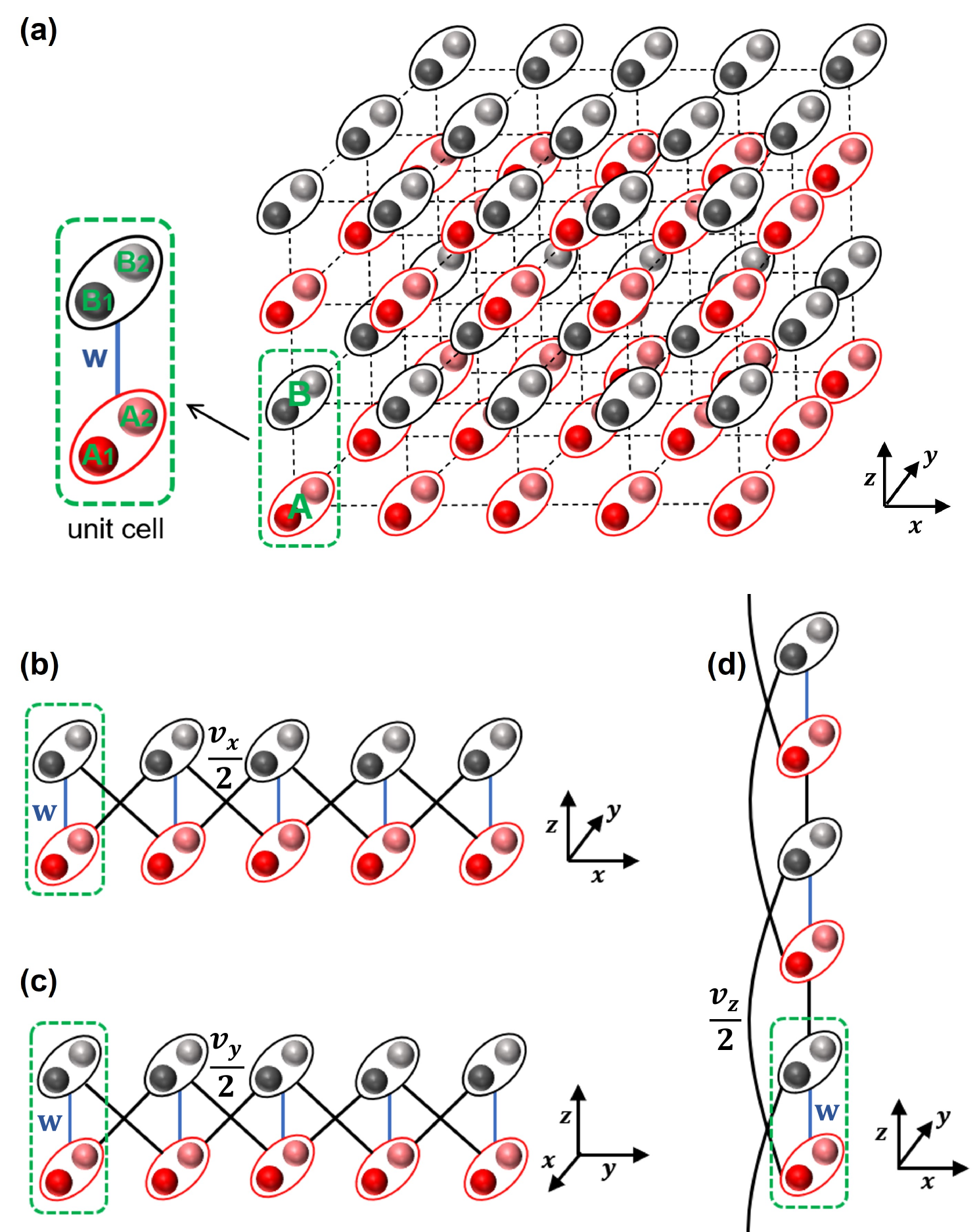}\caption{(a) is the corresponding lattice model of Eq. (\ref{eq:H3dDIII02}).
The red layer and grey layer stacked in the $z$ direction form the space
of matrices $\tau_{0,x,y,z}$. (b), (c) and (d) show the hoppings in the $x$-, $y$-, and $z$-direction.  \label{fig:ChiralTauzHoping}}
\end{figure}

\section{Surface states\label{ap:surface_states}}
Here we show how we obtained the surface states of the 3D models. 
Consider the 3D DIII model in Eq.~(\ref{eq:H3dDIII01}) as an example. When $w$ goes across $-3v$, topological phase transition happens with the bulk gap closing and reopening at $k_x,k_y,k_z=0$. 
Near the phase transition when $w\sim-3v$, the physics can be described by the low energy Hamiltonian:
\begin{align}
\label{eq: hDIII_lowen}
h_{\text{DIII}}\left(k_x,k_y,k_z\right)=&  
v \left(k_{x}\tau_{z}\sigma_{x}+k_{y}\tau_{z}\sigma_{y}+k_{z}\tau_{x}\sigma_{0}\right)\nonumber\\
&+\left(m-v k^2\right)\tau_{z}\sigma_{x},
\end{align}
where $k^2=k_x^2+k_y^2+k_z^2$  and the mass term $m=w+3v$. 
If there is an open boundary perpendicular to the $z$ direction at $z=0$, $k_z$ is not a good quantum number anymore,  however, the system remains periodic in $x-$ and $y-$ directions and thus is characterized by well-defined momentum components $k_x$ and $k_y$.  The corresponding Hamiltonian in the presence of boundary condition becomes (\ref{eq: hDIII_lowen}) with $k_z\rightarrow -ia\partial_{z}$, where $a$ is the lattice constant in $z$ direction. To find the two degenerate surface states at the center of the Dirac cone, set $k_x,k_y=0$,  so the three-dimensional Hamiltonian becomes a one-dimensional Hamiltonian in the $z$ direction.  Given the model is in the region $z>0$, the boundary condition requires that the wavefunctions exist at $z>0$ and vanish at $z=0$.  By solving the corresponding Schroedinger equation,  one can find there exist zero-energy bound states at the surface in the case $mv>0$ described by the wavefunctions:
\begin{equation}
\Psi_1\left(z\right)={C}\psi_{A}\left(e^{-z / \xi_{+}}-e^{-z / \xi_{-}}\right),\label{Psi1}
\end{equation}
\begin{equation}
\Psi_2\left(z\right)={C}\psi_{C}\left(e^{-z / \xi_{+}}-e^{-z / \xi_{-}}\right),\label{Psi2}
\end{equation}
where the penetration depth
$\xi_{\pm}^{-1}=\frac{1}{2a}\left(1\pm\sqrt{1-4m/v}\right)$
and $C$ is the normalization constant.  The eigen energies of $\Psi_1\left(z\right)$ and $\Psi_2\left(z\right)$ are both zero.
The eigen vectors $\psi_A=\left\{-i,0,0,1\right\}^T/\sqrt{2}$,  $\psi_C=\left\{0,-i,1,0\right\}^T/\sqrt{2}$, same as those of the surface states shown in  Fig. \ref{fig:Couple2dD}(a). 
Eq.~(\ref{Psi1}) and (\ref{Psi2}) describe the wavefunctions at $z\geq 0$. They both decay exponentially away from the boundary at $z=0$ to the region $z>0$. In the region $z<0$, we make $\Psi_{1,2}\left(z<0\right)=0$. Note we are now focusing on the case when $w$ goes across $-3v$ and $w$ satisfies $\left|w\right|<3\left|v\right|$. Then the bound state solutions exist as $mv>0$ and $1>\sqrt{1-4m/v}>0$.

The effective surface model can be obtained by projecting the bulk Hamiltonian onto the surface states $\Psi_1\left(z\right)$ and $\Psi_2\left(z\right)$ in Eq.~(\ref{Psi1},\ref{Psi2}).
This will lead to a $2\times 2$ effective Hamiltonian:
\begin{align}
h_{\text{DIII}}^{\text{sur}}&=\int dz U^{\dagger}\left(z\right)h_{\text{DIII}}\left(k_x,k_y,w -ia\partial_{z}\right)U\left(z\right)\nonumber\\
&=v\left(k_x \eta_y+k_y \eta_z\right),\label{eq:BoundaryEffkxky}
\end{align}
where $U\left(z\right)=\left\{\Psi_1\left(z\right),\Psi_2\left(z\right)\right\}$, $\eta_{x,y,z}$ are Pauli matrices. 

In the same way, one can derive the effective Hamiltonian of the surface states of the model in Eq.(\ref{eq:H3dDIII02}) with the same open boundary conditions. The surface states at $k_x,k_y=0$ are in the same form as those of (\ref{Psi1}) and (\ref{Psi2}). The corresponding  eigen vectors are $\psi_{\alpha}=\left\{ 1,1,0,0\right\} ^{T}/\sqrt{2},\psi_{\beta}=\left\{ 0,0,-1,1\right\} ^{T}/\sqrt{2}$, which are the eigen vectors we used in Sec. \ref{sec:h0surface} in the main text. The penetration depths are also $\xi_{\pm}$. Also, we can obtain the effective model Hamiltonian near $k_x,k_y=0$ by projecting the model Hamiltonian to the surface states:
\begin{equation}
h_{0}^{\text{sur}}=v\left(k_{x}\eta_{y}-k_{y}\eta_{x}\right),\label{eq:STofDIIIAppdix}
\end{equation}

In addition, the three symmetry operators of model~(\ref{eq:H3dDIII02}) $T_0$, $P_0$ and $C_0$ can also be projected onto the surface states because they preserve all of the three symmetry operators. The symmetry operators do not affect the $z$-dependent part of the wave functions and only act on the eigenvectors $\psi_{\alpha,\beta}$. Defining the projection matrix $U_P=\left(\psi_{\alpha},\psi_{\beta}\right)$, the three projected symmetry operators are $T_{0}^{\text{sur}}=U_P^{\dagger}T_{0}U_P$, $P_{0}^{\text{sur}}=U_P^{\dagger}P_{0}U_P$ and $C_{0}^{\text{sur}}=U_P^{\dagger}C_{0}U_P$. Next we will prove that $T_{0}^{\text{sur}}$, $P_{0}^{\text{sur}}$ and $C_{0}^{\text{sur}}$ are the symmetry oprators of effective surface Hamiltonian (\ref{eq:STofDIIIAppdix}). By introducing another two orthogonal vectors $\psi_{\gamma}$ and $\psi_{\eta}$ that are also orthogonal to $\psi_{\alpha,\beta}$,  together with $\psi_{\alpha}$, $\psi_{\beta}$  one has a complete set of vectors. Then one can obtain the form of symmetry operators in the new basis of $\psi_{\alpha - \delta}$ using the unitary matrix $U=\left(\psi_{\alpha},\psi_{\beta},\psi_{\gamma},\psi_{\delta}\right)$. 
Since the surface states preserve all of the symmetry operators, if any symmetry operator acts on an eigenvector of the surface states $\psi_{\alpha}$ or $\psi_{\beta}$,  the result will be a linear combination of $\psi_{\alpha}$ and $\psi_{\beta}$. This means that in the new basis, all the symmetry operators are block-diagonal. Take chiral symmetry $C_0$ as an example, 
\begin{equation}
U^{\dagger}C_{0}U=
\left(\begin{array}{cc}
C_{11} & 0\\
0 & C_{22}
\end{array}\right).
\end{equation}
The model preserves the chiral symmetry $ChC^{-1}=-h$, where $h$ is the $k_x$-, $k_y$- dependent Hamiltonian which is going to be projected to the surface state vectors (the $z$ dependent part has been integrated out). In the new basis, the relation becomes
\begin{equation}
\left(U^{\dagger}CU\right)\left(U^{\dagger}hU\right)(U^{\dagger}CU)^{-1}=-U^{\dagger}hU
\end{equation}
This relation can be further written as 
\begin{align}
&\left(\begin{array}{cc}
C_{11} & 0\\
0 & C_{22}
\end{array}\right)
\left(\begin{array}{cc}
h_{11} & h_{12}\\
h_{21} & h_{22}
\end{array}\right)
\left(\begin{array}{cc}
C_{11}^{-1} & 0\\
0 & C_{22}^{-1}
\end{array}\right)\nonumber\\
&=-\left(\begin{array}{cc}
h_{11} & h_{12}\\
h_{21} & h_{22}
\end{array}\right)
\end{align}
Note that $C_{11}=C_0^{\text{sur}}$ and $h_{11}=h_0^{\text{sur}}$, we have 
\begin{equation}
C_0^{\text{sur}}h_0^{\text{sur}}\left(C_0^{\text{sur}}\right)^{-1}=-h_0^{\text{sur}}
\end{equation}
So far we have proved that the projected operator $C_0^{\text{sur}}$ is the chiral symmetry operator of $h_0^{\text{sur}}$ in Eq.~(\ref{eq:STofDIIIAppdix}). In this way, one can also prove that $T_0^{\text{sur}}$ and $P_0^{\text{sur}}$ are the time-reversal symmetry operator and particle-hole symmetry operator of the effective surface Hamiltonian $h_0^{\text{sur}}$.

\bibliographystyle{apsrev4-1} 
\bibliography{3Dchiral}

\end{document}